\def\apjournal{0}

\ifnum\apjournal =1
\documentclass[12pt,preprint]{aastex}
\else
\documentclass{emulateapj}
\fi

\usepackage{color}
\usepackage{natbib}
\usepackage[geometry]{ifsym}
\usepackage{amsmath} 

\newcommand{\vect}[1]{\textbf{#1}}
\newcommand{\ddt}[1]{\frac{\partial #1}{\partial t}}

\begin{document}

\ifnum\apjournal =0
\submitted{Received 2009 May 29; accepted 2009 August 26}
\fi

\title{Photospheric Motions and Their Effects on the Corona: \\
    a Numerical Approach}

\ifnum\apjournal =0
\author{L. F. Gomes de Jesus and B. V. Gudiksen}
\affil{Institute of Theoretical Astrophysics, University of Oslo, P.O. Box 1029 Blindern, N-0315 Oslo, Norway}
\else
\author{Leandro Filipe Gomes de Jesus}
\email{leandro@astro.uio.no}
\and
\author{Boris Vilhelm Gudiksen} 
\email{b.v.gudiksen@astro.uio.no}
\affil{Institute of Theoretical Astrophysics, University of Oslo, P.O. Box 1029 Blindern, N-0315 Oslo, Norway}
\fi

\begin{abstract}
We perform a number of numerical simulations of the solar corona with the aim to understand how it  responds  to different conditions in the photosphere. By changing parameters which govern the motion of the plasma at the photosphere we study the behavior of the corona, in particular, the effects on the current density generated. An MHD code is used to run simulations, using a $20\times20\times20~\rm{Mm^3}$ box with time spans ranging from one hundred to several hundreds of minutes. All the experiments show a fast initial increase of the current density, followed by a stabilization around an asymptotic value which depends on the photospheric conditions. These asymptotic average current densities as well as the turn-over points are discussed.
\end{abstract}


\section{Introduction}

Over the last decades the solar corona has been, and continues to be scrutinized and studied in a search for the explanation of its heating mechanism \citep{rosner78,golub80,hendrix96,aschwanden01}. With the advent of the computer, simulations become feasible and the reproduction of observations through models have advanced our understanding of the 1 MK solar corona \citep{gudiksen05b}.
Among the theoretical work, two main ideas have become popular to explain the hot corona. One supporting a wave heating mechanism introduced by \citet{alfven47} where the energy is carried through magneto-hydrodynamic waves; and the other based on electric currents developed by the movements of the magnetic field lines, first suggested by \citet{parker72}. 

It has been clear from the beginning that the magnetic field play a very important role in the process of coronal heating, besides passively shaping coronal structures. It works as an active agent carrying energy from the convective outer layer of the Sun, and participates in the energy storage and dissipation mechanism. The continuous convective motions shuffle the footpoints of the magnetic field lines, twisting the magnetic field lines and increasing the complexity of the magnetic field configuration. In the context of the Parker nanoflare model, the kinetic energy is transmitted to the fluid and converted into heat by a dynamical non-equilibrium and, in spite of the very high electrical conductivity of the plasma, the corona is heated through the dissipation of current sheets. Several studies and observations have been made supporting these ideas, associating the coronal heating with the stress of magnetic field by the surface turbulence and showing a direct relation between the coronal emission and the location of the magnetic fields on the surface (\citealt{vanSpeybroeck70,vaiana73,krieger71,krieger76} to mention examples).

It is known that main-sequence stars with masses $M<1.3~M_{\Sun}$ (spectral types F and later) have a nuclear burning core, surrounded by a shell with radiative transport and an outer convective envelope where the energy is carried out of the star by convective motions. The convection zone is narrow in extent in F-type stars, but in the Sun (G2 V) it has a depth of 0.287 solar radii \citep{christensen91} and, with decreasing effective temperature the convective zone occupies a larger and larger fraction of the stellar volume with late M stars being fully convective. The photosphere lies above this convective layer and marks a transition between a medium where energy is carried by convection to a radiative efficient domain. The granular structure at the stellar surface is fundamentally different from the flow topology below the photosphere. Numerical models suggest that convection is mostly driven by the downflows due to radiative cooling near the surface \citep{stein98} and that these downflows converge into downdrafts, forming tree-like structures down into the convection zone, where the intergranular lanes combine into more and more widely separated branches \citep{spruit90}\footnote{See page 291, Figure 7 for an illustration of this scenario.}.

As a result of upflows and downflows of plasma, the evolution of the cellular network in the photosphere is quite complex with granules growing, decaying and merging with other granules \citep{spruit90}, but some studies have been carried out trying to define and model some physical characteristics of solar granulation like the size of granules (for example, \citealt{roudier87,schrijver97}), lifetimes \citep{spruit90} and flow profiles of the plasma (see \citealt{canfield76} and \citealt{durrant79} for estimates of vertical velocity component, and \citealt{beckers70} for the horizontal velocity component). The differences between stellar convection zones (see \citealt{nordlund90} for some simulations regarding the convection in some solar-like stars), makes it likely that the structure of the granulation on other stars differs from the granulation pattern we have on the Sun. The change in magnitude and topology of the granular velocity field should affect the total Poynting flux transmitted into the corona, and most likely the amount of heat dissipated in stellar atmospheres.

Using a numerical code to simulate the granular motions at the photosphere, we follow the effects on the corona by measuring the electric current density. Running the code for several values of parameters associated with the structure of the granular motions, we investigate what happens with the corona as a function of the granular characteristics.

\section{Numerical Procedure}

\subsection{Method}\label{sec:method}

The simulations are carried out with the 3-D MHD code extensively described in \citet{gudiksen05b}, which solves the fully compressible MHD equations:

\begin{eqnarray}
\ddt{\rho} &=& -\nabla \cdot \rho \vect{u}, \label{continuity}\\
\ddt{\rho \vect{u}} &=& - \nabla \cdot (\rho \vect{u} \vect{u} - {\underline{\underline{\tau}}}) - \nabla P + \vect{J} \times \vect{B} + \rho\vect{g}, \label{moment}\\
\ddt{e} &=& -\nabla \cdot (e \vect{u}) - P \nabla \cdot \vect{u} + Q_{Joule} + Q_{visc}, \label{energy}\\
\ddt{\vect{B}} &=& -\nabla \times \vect{E}\\
\vect{E} &=& \eta \vect{J} - \vect{u} \times \vect{B},\\
\mu \vect{J} &=& \nabla \times \vect{B}.
\end{eqnarray}

Here $\rho$, $\vect{u}$, $P$, $\vect{J}$, $\vect{B}$, $\vect{g}$, $\mu$, $\vect{E}$ and $e$ denote the mass density, velocity vector, gas pressure, electric current density, magnetic flux density, gravitational acceleration, the vacuum permeability, electric field vector and the internal energy per unit volume respectively; The terms ${\underline{\underline{\tau}}}$ and $\eta$ refer to the viscous stress tensor and the magnetic diffusivity; $Q_{visc}$ and $Q_{Joule}$ represent the viscous and Joule heating.

The set of MHD equations is computed on a staggered mesh of $100^3$ grid points corresponding to a box of $20\times20\times20~\rm{Mm^3}$ (distance between two grid points of $0.2~\rm{Mm}$), employing 6th-order derivative operators, 5th-order interpolation operators to align physical variables in space, and a 3rd-order Runge-Kutta time-stepping scheme.

Simulations on this work use a classical heat conduction along the magnetic field, Spitzer conductivity \citep{spitzer56} and also a cooling function representing radiative losses in the optically thin coronal plasma.
\citet{parker72} argued that the vigorous formation of tangential discontinuities (current sheets), magnetic dissipation is practically independent of the conductivity. Moreover, it was showed numerically by \citet{galsgaard96} and also \citet{hendrix96} that the energy created through dissipation is not strongly dependent on small scales. One should expect that, by resolving the smallest scales in a simulation of DC heating, the dissipated energy could have a different value. Instead, high resolution does not give dramatic changes in results as the stress applied on the magnetic field is not increased but rather shifted to a fine scale. The important point on the ohmic/nanoflare heating hypothesis is the level of stress experienced by the magnetic field which does not depend on the resolution scale, so consistent results can be retrieved from simulations without the necessity of using very high resolution.

The MHD equations are solved in a 3D x-y-z coordinate system with periodic boundary conditions in the horizontal direction.
The top of the box has zero-gradient of temperature at the boundary so no heat exchanges are allowed. Also the vertical gradient of the horizontal velocity is null at the upper boundary. Both the density and potential field are extrapolated in the ghost zones (regions outside the computational domain). Vertical velocity is set to zero on both top and bottom boundaries and the driver establishes the horizontal velocity at the bottom. The lower boundary is stressed by a time dependent velocity field, constructed from a Voronoi tesselation, reproducing the granulation pattern. The velocity field reproduces the geometrical pattern as well a the amplitude power spectrum of the velocity and the vorticity, leaving no free parameters.

The photospheric motions are produced through a Voronoi tessellation driver \citep{okabe00} which reproduce the geometry and flow pattern of the granular surface of the Sun. A particular model of a Voronoi tesselation investigated by \citet{schrijver97}, the so-called ``multiplicatively weighted Voronoi'' tessellation, was found to reproduce the granular pattern of the Sun very well. It segments the two-dimensional space into Voronoi regions $V(\vect{p}_{i})$ defined by
\ifnum\apjournal =1
\begin{equation}
V(\vect{p}_i ) = \left\{ {\vect{x}\left| {\frac{{w_i }}{{\left\| {\vect{x} - \vect{x}_i } \right\|^p }} \ge \frac{{w_j }}{{\left\| {\vect{x} - \vect{x}_j } \right\|^p }},~~{\rm for}~~j \ne i,j \in \left[ {1,...,n} \right]} \right.} \right\} \label{eq:vor-tess}
\end{equation}
\else
\begin{eqnarray}
V(\vect{p}_i )  &= & \Biggr\{ {\vect{x} \left| {\frac{w_i}{{\left\| {\vect{x} - \vect{x}_i } \right\|^p }}} \ge \frac{{w_j }}{{\left\| {\vect{x} - \vect{x}_j } \right\|^p }} \right.}, \nonumber\\
& & ~~~~~~~~~~~~~~~~~~~~{\rm for}~~j \ne i,j \in \left[ {1,...,n} \right] \Biggr\} \label{eq:vor-tess}
\end{eqnarray}
\fi

with $w_{i}$ representing the different weights (greater than zero) associated with the generator points $\vect{x}_{i}$, and $p$ denoting the index of the power law function which controls the tessellation. By using these weighted distances between the generated points, the generation of very large or very small granules can be avoided. By choosing the weights $w_i$ carefully, a uniform granular structure is produced with only acceptable sizes for the granules, ranging in size between 1 and 2 Mm, in the case of solar photospheric granules, to $\sim$ 30 Mm or larger for supergranules \citep{leighton63}.

The velocity profile used by the model is taken from \citet{simon89}, who fitted observational data from the SOUP instrument on \textit{Spacelab 2}, and then adjusted to better fit the solar velocity spectrum by
\begin{equation}
v_h (r) = v_{i}(t) r^2 e^{ - \left( {\frac{r}{{r_0 }}} \right)^2 } \label{eq:velprofile}
\end{equation}
In the previous equation $v_{i}$ is a time-dependent quantity, proportional to $w_{i}(t)$ which is defined by
\begin{equation}
w_i (t) = W_i \exp \left[ { - \left( {\frac{{t - t_{0,i} }}{{\tau _i }}} \right)^q } \right]
\end{equation}
with $W_{i}$ being the maximum weight and $q$ an even integer related with the growth time of the granule. The velocity at the center of granules evolve in time along a Gaussian profile $ v_i \propto e^{-(t/t_0)^2}$.

\subsection{Simulations}

\ifnum\apjournal =0
\begin{deluxetable*}{ccccccccc}[b]
\tablecaption{Main parameters for the simulations \label{tab-1}}
\tablehead{
\colhead{} & \colhead{} & \multicolumn{3}{c}{Input parameters} & \colhead{} & \colhead{} & \colhead{}\\
\cline{3-5}
\colhead{Simulation} & \colhead{Set} & \colhead{$r$ (Mm)} & \colhead{$\tau$ (min)} & \colhead{$v~\rm{(km \cdot s^{-1}})$} & \colhead{\# Snapshots} & \colhead{$t$ (min)}
}
\startdata
I & 1, 2, 3 & 2 & 5 & 3 & 600 &  259.6\\
II & 1 & 3 & 5 & 3 & 600 & 251.3\\
III & 1 & 5 & 5 & 3 & 300 & 128.7\\
IV & 2 & 2 & 8 & 3 & 750 & 318.7\\
V & 2 & 2 & 12 & 3 & 1000 & 414.1\\
VI & 3 & 2 & 5 & 6 & 600 & 261.3\\
VII & 3 & 2 & 5 & 10 & 750 & 326.8\\
\enddata
\tablecomments{Variables on the table are $r$, the size of the granules; $\tau$, the lifetime of the granules; $v_{0}$, the plasma flow velocity, $t$, the time-span of the simualtion.}
\end{deluxetable*}
\fi

To quantify the effect of the photospheric granular network on the corona, we set up three sets of simulations. In each set we vary one of three parameters describing the granules: The granular radius $r$, the granular lifetime $\tau$ and the velocity amplitude of the plasma $v_0$. Each set contains three simulations with one simulation being identical for all the sets so the total number of simulations is 7. The details of the simulations and sets are provided in Table~\ref{tab-1}. 

Simulation I is common to all the three sets and corresponds to the closest reproduction of solar conditions, by choosing $r=2~\rm{Mm}$, $\tau=5~\rm{min}$ and $v_{0}=3~\rm{km~s^{-1}}$. It is important here to point out that these parameters are not entirely independent and setting these values does not produce granules that have exactly these characteristics. The actual values for the granular radius, granular lifetime and average velocity in the granular field will be discussed in Sec. \ref{sec:vel-field}.

In set 1 both the lifetime and velocity amplitude are kept constant, while the granular radius is varied from 2 to 5 Mm. In the second set we simulate granules with lifetimes of 5, 8 and 12 minutes, maintaining the size of the granules and velocity amplitude constant and equal to 2 Mm and $3~~\rm{km~s^{-1}}$ respectively. The third set includes simulations with three different flow velocities, 3, 6 and 10 $\rm{km~s^{-1}}$ and granule sizes and lifetimes set to 2 Mm and 5 minutes, respectively. 

Our simulation domain is initially a current-free space containing vertical magnetic field lines extending from the base which coincides with the photosphere ($z=0$) to the top of the box, 20 Mm up in the corona ($z=H=20~\rm{Mm}$). We then let the code run for several hundreds of minutes. The temporal resolution, i.e. the interval between two consecutive snapshots produced by the model is roughly half a second, thus providing us with an almost continuous evolution of the desired variables. We should note that the time span of each simulation is dependent on how fast the electric current density reaches a steady-state. After this point (designated here as the turn-over point) we extend the simulation, in order to ensure that the steady-state was reached. 

\section{Results}

We are only interested in the behaviour of the dissipated current in the corona. For all the simulations we follow the average current density in the corona because it is a direct measure of the total dissipated magnetic energy. To avoid effects of the timevarying local height of the transition region, we calculate the average current density, taking into consideration the volume between $z=5~\rm{Mm}$ and $z=15~\rm{Mm}$.
The size of our box is small compared with the typical size of  coronal loops. The purpose is to study a portion of the loop from just above the transition region up to several mega-meters into the corona. This will correspond to one of the loop legs where the heating becomes more important. The magnetic field becomes nearly force-free above the transition region and the current along each field line will be proportional to the magnetic field and as described by \citet{gudiksen05a} the average heating (which is proportional to the magnetic field strength squared) drops exponentially above that layer. 

One behavior is common to all the simulations. The evolution of the average current density in the corona shows a very steep increase at the beginning of each simulation, followed by a gradual deceleration and a turn-over point, and later a stabilization around a constant level designated here as asymptotic average current density squared ($J_A^2$). Nonetheless, some differences exist between the experiments.

To better identify and understand these differences we perform a curve fit for the data assuming a special variation of the logistic function, given by
\begin{equation}
\label{eq:fit}
F(t) = a_0 \cdot e^{ - \frac{{a_1 }}{t}} 
\end{equation}
where $a_{i},  i=0,1$ are the two free fit model parameters.  One important property of this function is that it has a horizontal asymptote, i.e. $F_{asym} = \mathop {\lim }\limits_{t \to \infty } F(t) = a_0$, so it can be used to determine the asymptotic level of the average current density when the box reaches a steady-state. The turn-over point ($t_{turn}$) is defined here as the time when the fitting-function reaches 80\% of the asymptotic value of the fitting-curve:
\begin{equation}
F_{turn}=0.8 \cdot F_{asym}
\end{equation}

The turn-over points as well as the $J_A^2$ and the fitting parameters are summarized in Table~\ref{tab-2}.
In particular, results from the simulations are plotted in Figures~\ref{fig-1} -- \ref{fig-67}. The figures clearly  illustrat the fast initial growth of the average current density, strongly impelled by the sudden advection of the magnetic field lines. The random character of the movements of the footpoints is the natural way to interpret the fluctuations of the average current density even after the turn-over point.

\ifnum\apjournal =0
\begin{deluxetable}{cccccc}[b]
\tablecaption{Fitting parameters ($a_{0}$ and $a_{1}$) in Eq. \ref{eq:fit}, their uncertainties ($\epsilon_{a_0}$ and $\epsilon_{a_1}$) and turn-over points. \label{tab-2}}
\tablehead{
\colhead{} & \colhead{} & \multicolumn{4}{c}{Fitting-function}\\
\cline{3-6}
\colhead{Simulation} & \colhead{Turn-over point (min)} & \colhead{$a_{0}$} & \colhead{$\epsilon_{a_0}$} & \colhead{$a_{1}$} & \colhead{$\epsilon_{a_1}$}
}
\startdata
I & 54.5 & 6.23 & 0.06 & 12.2 & 0.7\\
II & 36.7 & 7.93 & 0.08 & 8.15 & 0.61\\
III & 25.7 & 11.6 & 0.2 & 5.78 & 0.49\\
IV & 95.4 & 8.76 & 0.11 & 21.2 & 1.3\\
V & 150.0 & 10.97 & 0.15 & 33.2 & 2.0\\
VI & 61.5 & 6.80 & 0.06 & 13.8 & 0.7\\
VII & 81.5 & 7.50 & 0.08 & 18.2 & 1.0\\
\enddata
\end{deluxetable}
\fi

Even though the granular pattern is changing, the statistical properties of the velocity field are constant, so the sharp rise in the average current must mean that the magnetic field is initially able to contain the energy injected into it by the Poynting flux produced by the photospheric velocity field. At the turn-over point a saturation is reached, and the magnetic field dissipates as much energy as is injected into it by the Poynting flux at the photosphere.

At first sight, the evolution of the average current density in the corona is quite similar for all the experiments, despite the variations of $r$, $\tau$ and $v_{0}$. Thus, if we do not consider the numbers involved but concentrate only on the behavior of the average current density, we conclude that it is practically impossible to distinguish between different simulations. After the exponential increase of the average current density, the magnetic field line mapping inside the box starts to become quite complicated and the shear pattern will evolve independently of the parameters which govern the motions of photospheric material. A similar argumentation is presented in \citet{galsgaard96} with respect to the variation of the driving boundary pattern.
\ifnum\apjournal =0
\begin{figure}[htb]
\epsscale{1.0}
\plotone{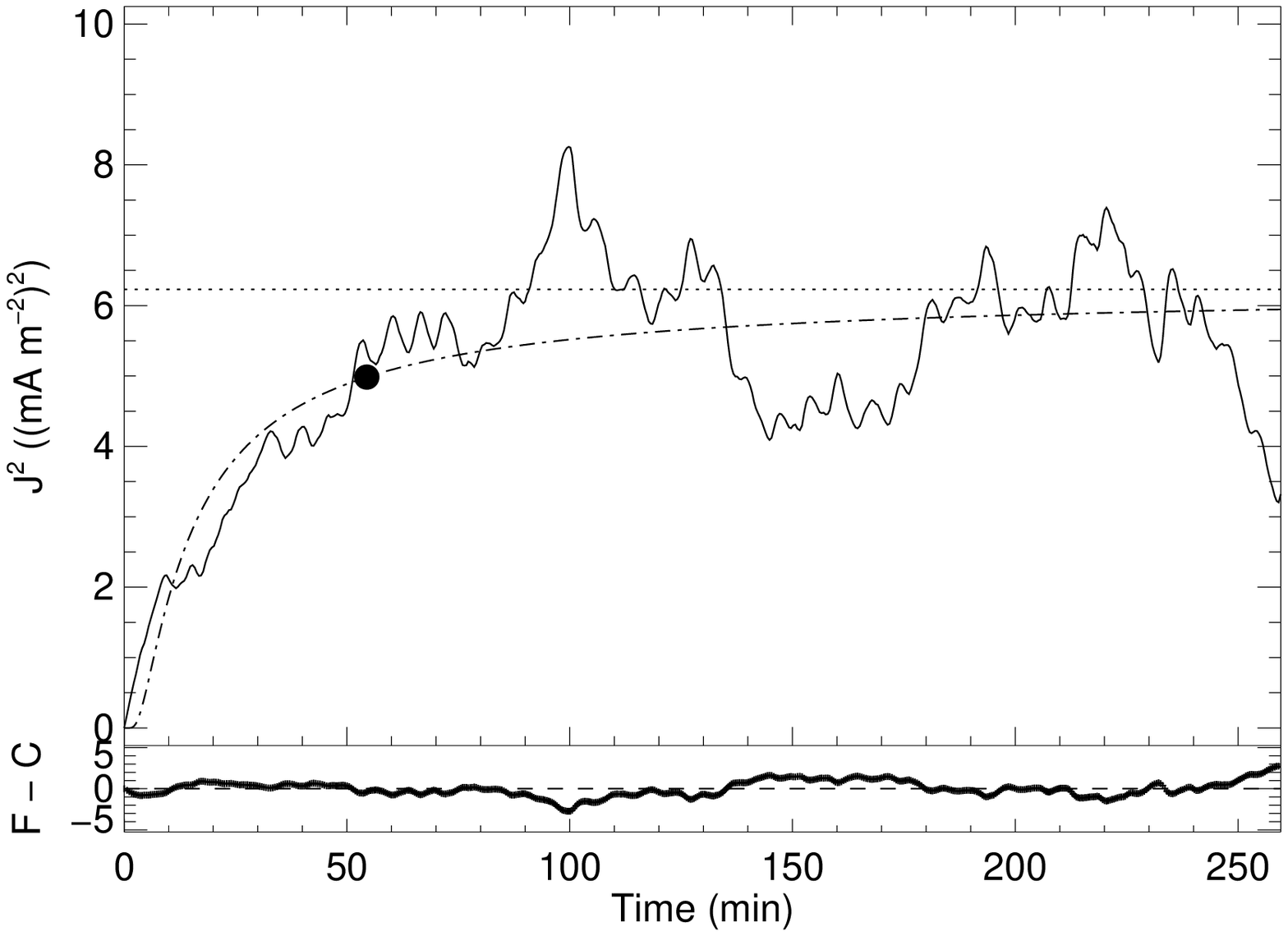}
\caption{Evolution of the average current density in the box as a function of time (solid line) and its curve fitting function (dash dotted line) in simulation I with granules size/ granules lifetime/ flow velocity = 2 Mm/ 5 min/ 3 $\rm{km~s^{-1}}$. Also shown are the $J_A^2$ (horizontal dotted line) and the turn-over point marked with the small filled circle over the fitting curve. The bottom part of the plot shows the residuals in the sense of (F-C: Fitted-Computed).}
\label{fig-1}
\end{figure}
\fi

\ifnum\apjournal =0
\begin{figure*}
\epsscale{1.0}
\plottwo{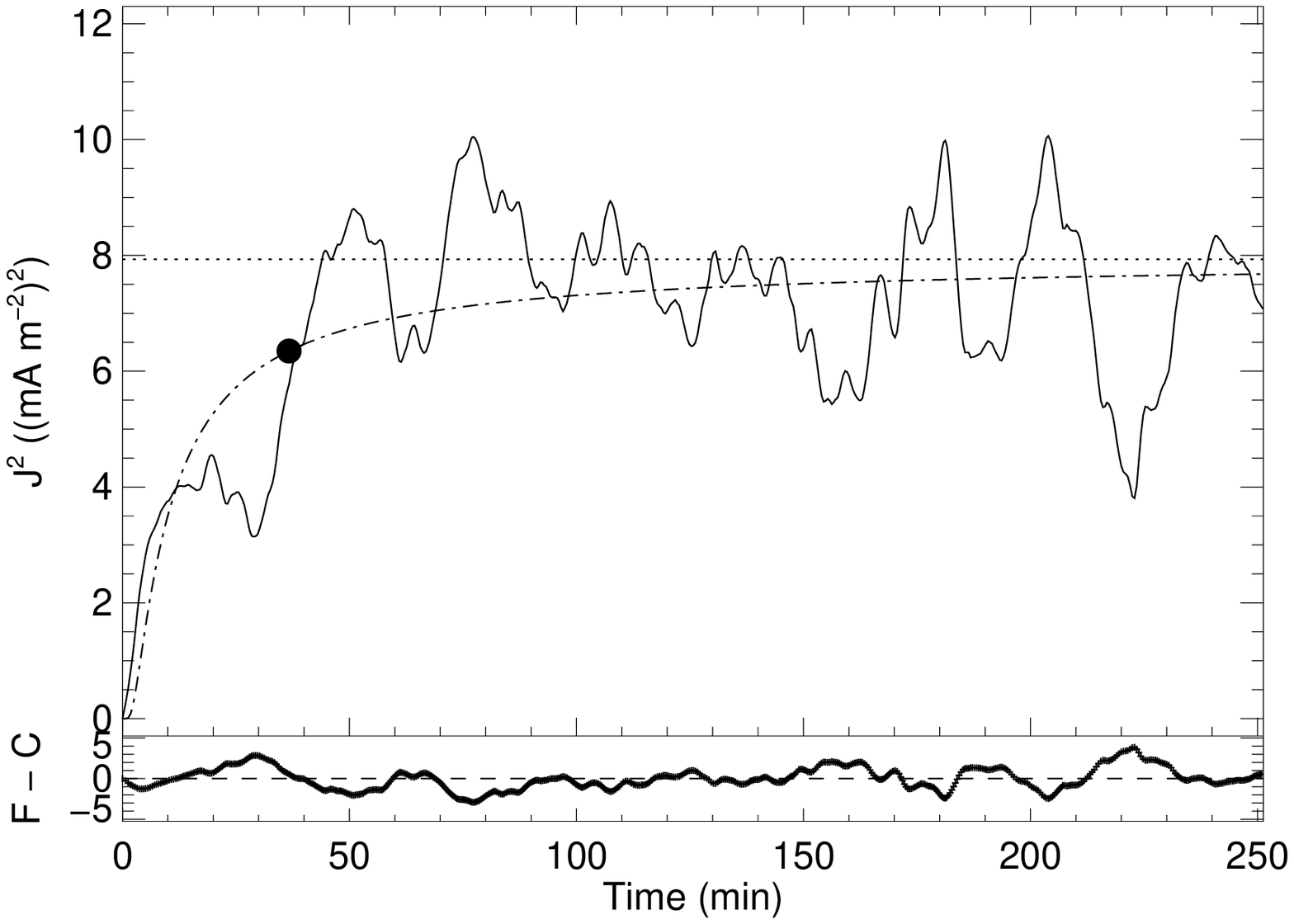}{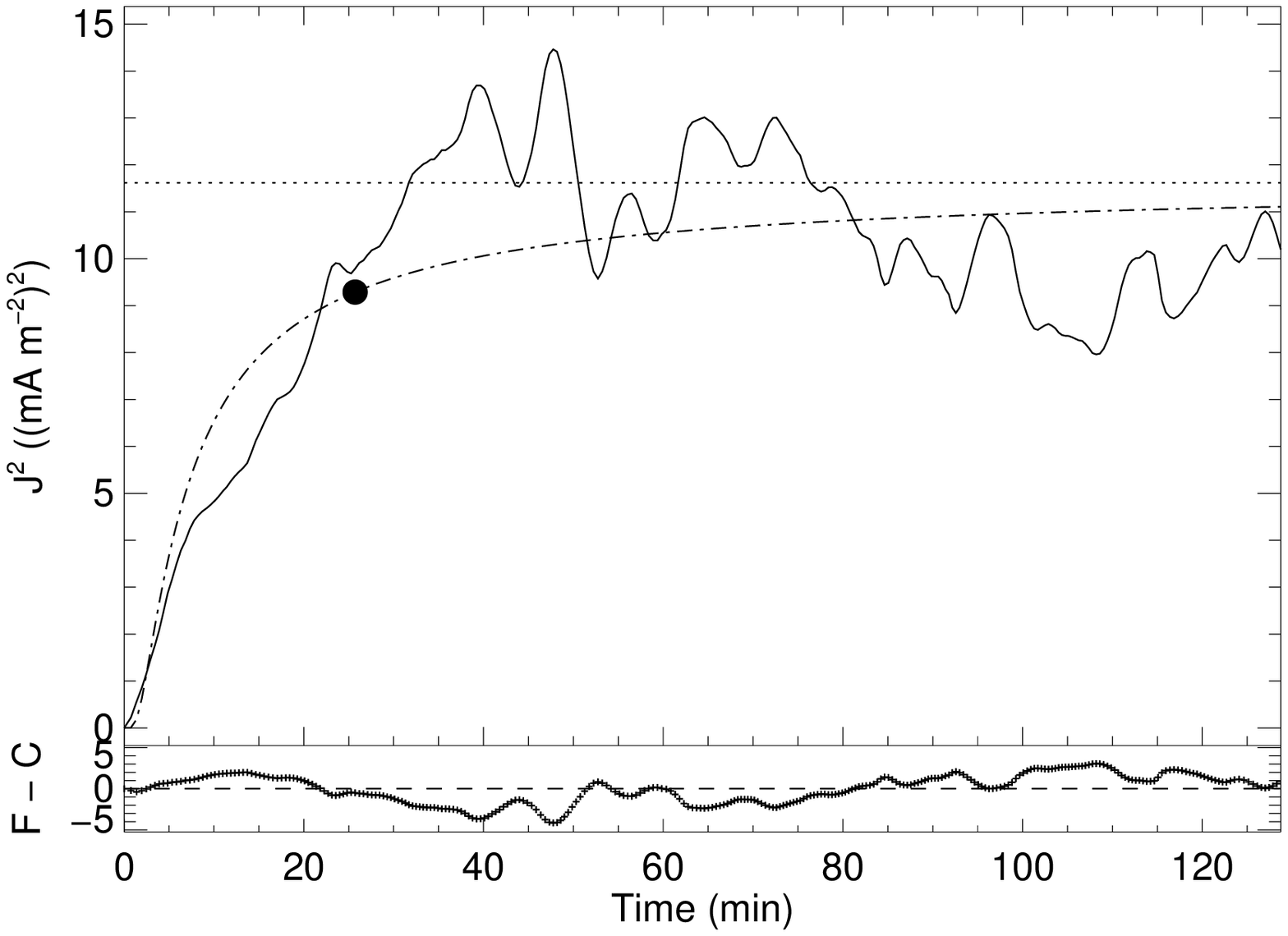}
\caption{Evolution of the average current density in the box as a function of time (solid line) and its curve fitting function (dash dotted line), in simulations II (left) and III (right).}
\label{fig-23}
\end{figure*}

\begin{figure*}
\epsscale{1.0}
\plottwo{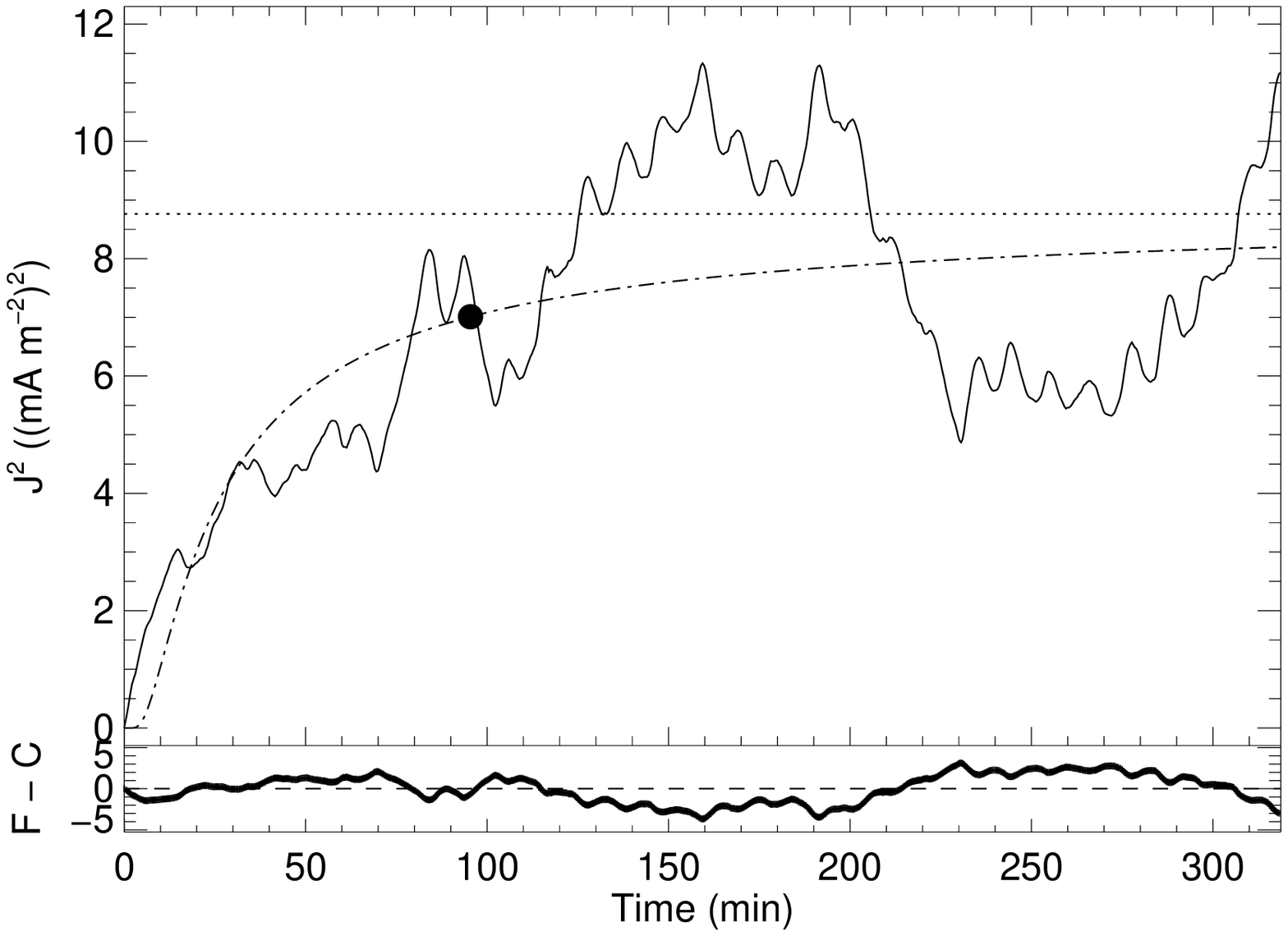}{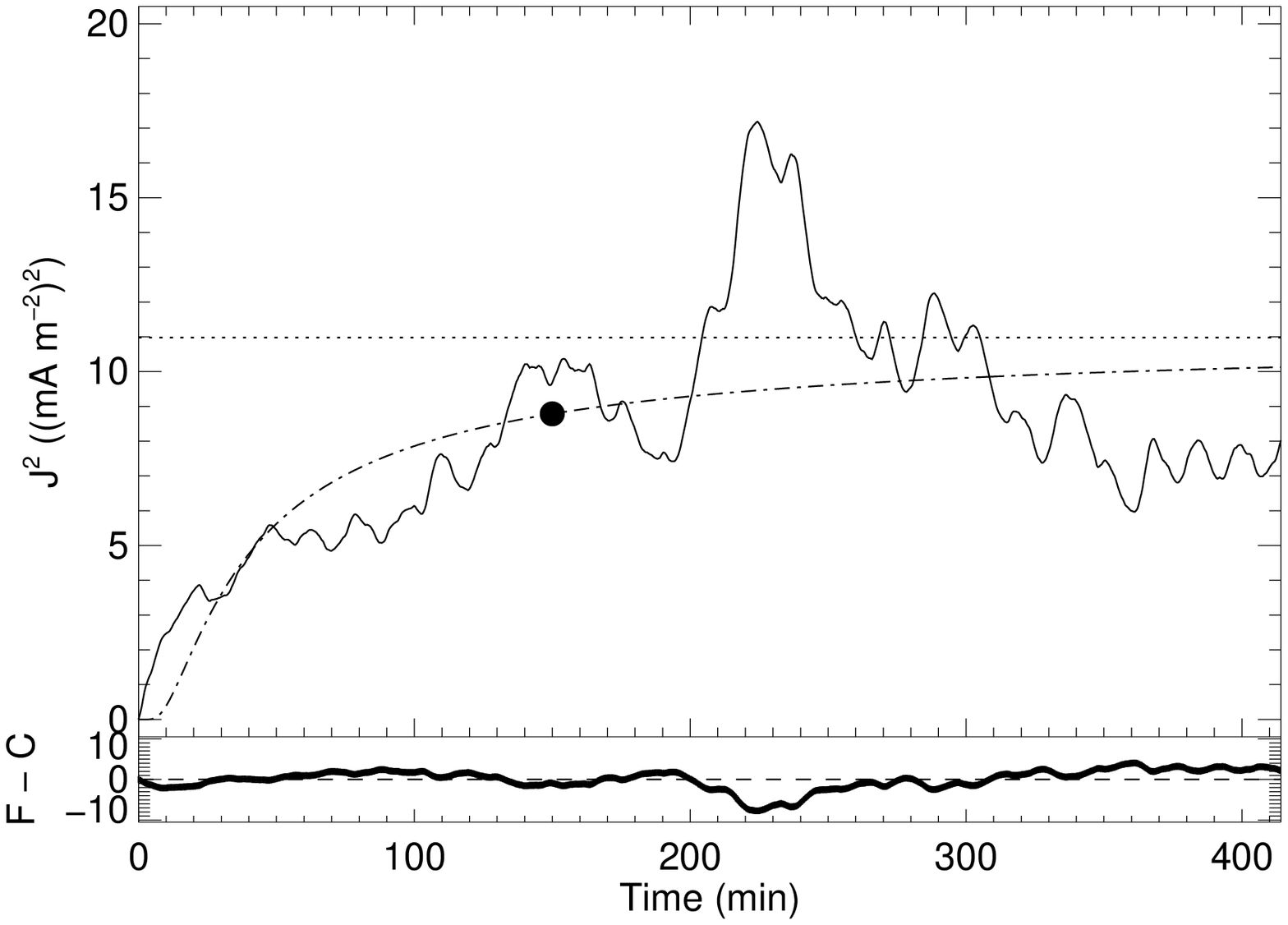}
\caption{Evolution of the average current density in the box as a function of time (solid line) and its curve fitting function (dash dotted line), in simulations IV (left) and V (right).}
\label{fig-45}
\end{figure*}

\begin{figure*}
\epsscale{1.0}
\plottwo{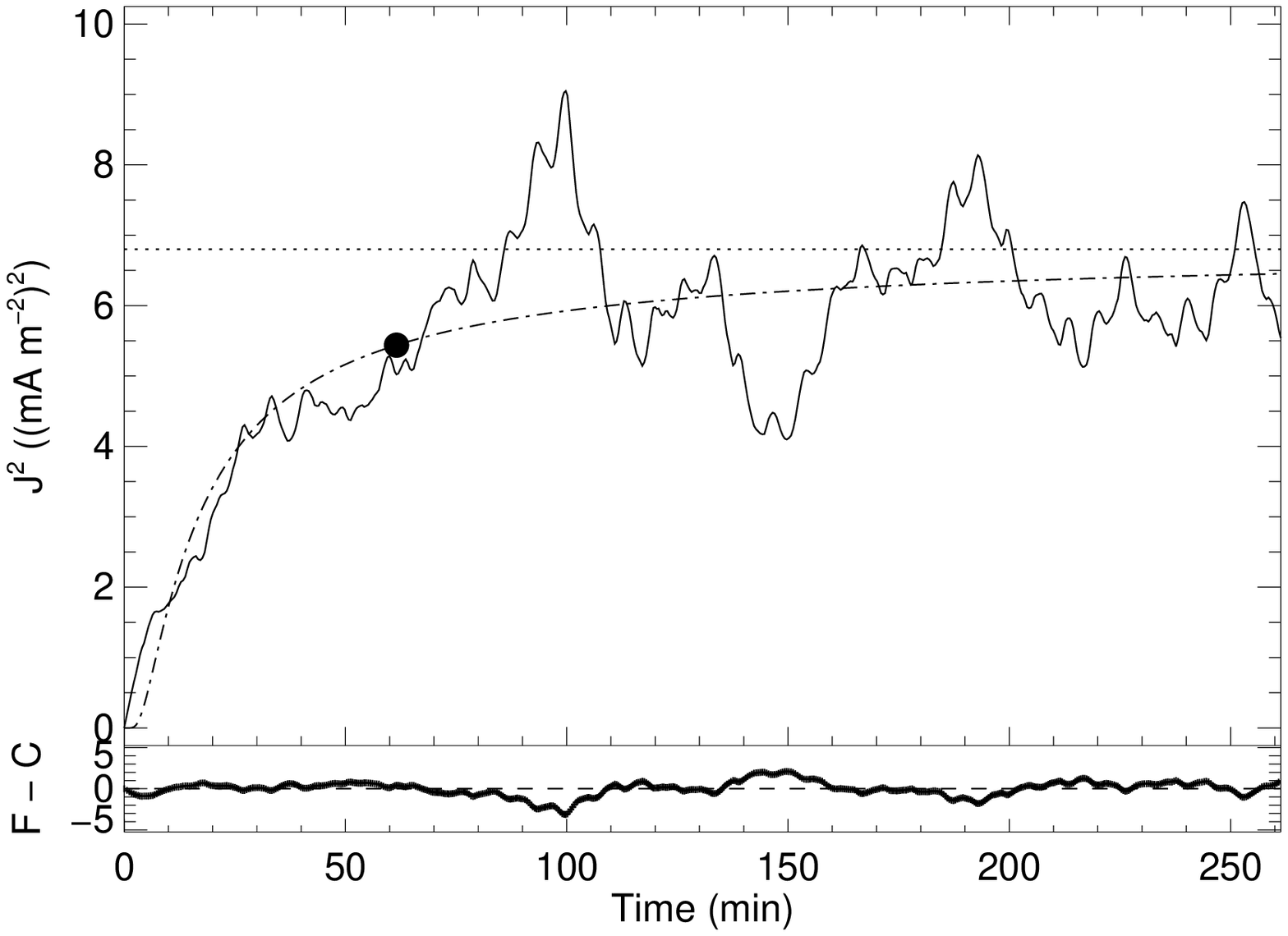}{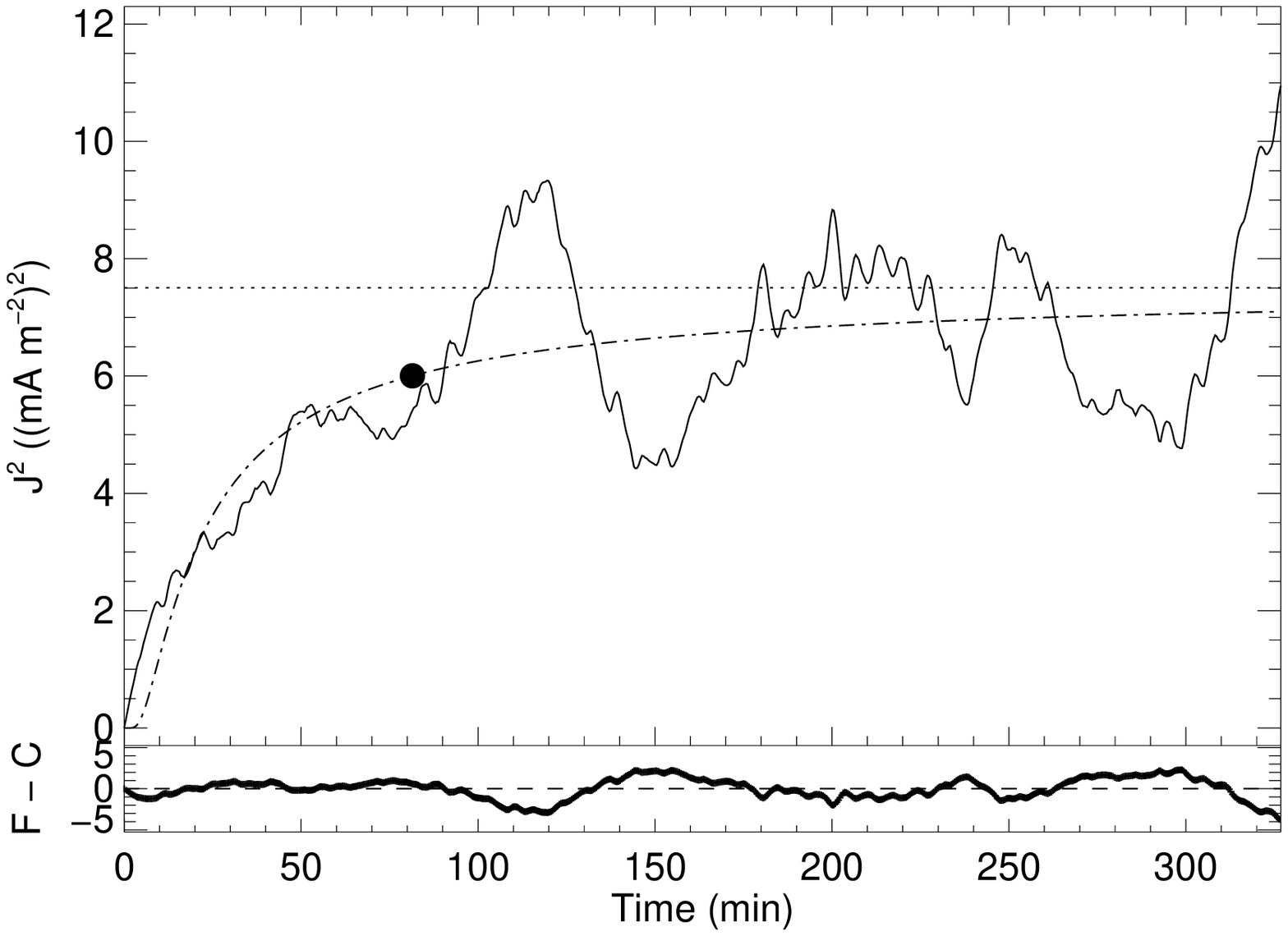}
\caption{Evolution of the average current density in the box as a function of time (solid line) and its curve fitting function (dash dotted line), in simulations VI (left) and VII (right).}
\label{fig-67}
\end{figure*}
\fi

\section{Discussion}
To better perceive the consequences of different photospheric configurations on the average current density, we analyze the moment of turn-over point and the $J_A^2$ at the steady-state as we change the three variables: granular size, granular lifetime and flow velocity of the plasma. This is done by having in the background the theory proposed by Parker for the coronal heating, based on the dissipation of energy through electric currents \citep{parker72,parker83a,parker83b,parker88}. Assuming a similar box with vertical magnetic field at $t=0$, he argued that the photospheric motions due to granulation drag the footpoints of the magnetic field, stressing it and creating current sheets which will dissipate and heat the corona.

Taking the velocity of the photospheric motions (equivalent to the shuffling velocity of the magnetic field lines) as $v$, the angle of deviation of the magnetic field as $\theta$, time as $t$ and vertical size of the box represented by $H$, the following equation,
\begin{equation}
\label{eq:theta}
\theta(t)  \approx \frac{{vt}}{H}
\end{equation}
is valid for relatively small values of $\theta$. By inserting the the vertical component of the magnetic field $B$, and the transverse component, $B_{\perp}=B~tan~\theta(t)$ we find that  
\begin{equation}
\label{eq:transverseB}
B_{\perp}  \approx \frac{Bvt}{H}.
\end{equation}
The onward motion of the footpoint creates tension in the field line in the opposite direction, so the movement of the footpoint does work on the magnetic field and the energy input grows as the field is dragged. After some time when $\theta$ reaches a ``critical angle'', the dissipation of the stressed field exactly counters the production of stress by the continuing motion of the footpoints. \citeauthor{parker83a} estimated the critical angle based on observation and energy arguments to be 20$^{\circ}$ \citep{parker83a} or 14$^{\circ}$ \citep{parker88}.

\subsection{The driving velocity field}\label{sec:vel-field}
The details of the driving velocity field in the photosphere is important. As seen, all the three parameters controlling the velocity field have an impact on the $J_A^2$ and the turn over point. 

In a simplified picture, the granular velocity field will advect magnetic footpoints radially in each granule, until they arrive at the intergranular boundary. Depending on the location of the surrounding granules, the magnetic footpoints will then either stay at the location in the intergranular lane or be advected almost perpendicular to their previous velocity until they reach an intersection between more than two granules, where the velocity is generally low. The development of new granules will make that process repeat itself. The direction of movement will be independent of the previous history of the magnetic footpoints, so they will experience a random walk, where the step length will be roughly the granular radius, unless the velocity of the flow inside the granules is so low that they will never have time to reach one granular radius before the granule disappears. In that case the random walk step length will be the product of the velocity inside the granule and the lifetime of the granule. A convincing argument that the magnetic footpoints do experience a random walk, is found by following magnetic footpoints in the granular velocity field, assuming they are passively advected (see Fig. \ref{fig-8} for an example of the path travelled by one footpoint). We have followed 1024 footpoints at the surface (forming a $32\times32$ grid at the bottom of the box) over time for each simulation. We then plot the displacement from the original position ($R$) as a function of time. Figure \ref{fig-9} shows that total displacement of the footpoints in the granular field of simulation I, and a fit of the form
\begin{equation}
\label{eq:randomwalk}
R = C t^{p} .
\end{equation}

In the previous expression $C$ and $p$ are constants and a value of $p = 0.5$ would correspond to theoretical random walk. The fitting curves have values of $p$ as indicated on Table~\ref{tab-3} which are fairly close to the theoretical value. We will assume that the motion experienced by a magnetic footpoint is a random walk, and we can then fit the data with
\begin{equation}
\label{eq:randomtheory}
R = \ell \sqrt{N} .
\end{equation}

The random walk is described by two parameters: the number of steps $N$ and the step length  $\ell$. The two parameters cannot be found independently, so they are not easy to find as we have only snapshots in time of the simulation. They can be found indirectly if we can find the step length $\ell$ and the step time $\bar \tau$ (the time required for one step). If these can be found the number of steps can also be determined since $N=t/\bar \tau$ where $t$ is the elapsed time. Because we have decoupled the velocity inside the granules, their lifetimes and their size, we have to be careful with the estimation of the parameters. 
The step length is assumed to be the smallest of these two values: the product of velocity times lifetime or the average radius of the granules at the surface, i.e., the distance from the center of each granule to the intergranular lanes/ vertices. By making this assumption we assume that the footpoint starts its movement always at the center of a granular cell which is a strong simplification and it will be discussed later. A footpoint can initiate the motion anywhere inside the granule migrating then towards the intergranular lanes, probably lying there for a while until a new granule emerges and propels it again. So, the granular radius should be understood as an over-estimated step length and the maximum value that this parameter can take. 

\ifnum\apjournal =0
\begin{deluxetable}{cccccc}[h]
\tablecaption{Values of $p$ and uncertainty $\epsilon_{p}$ in Eq. \ref{eq:randomwalk} for each simulation \label{tab-3}}
\tablehead{
\colhead{Simulation} & \colhead{p} & \colhead{$\epsilon_{p}$}
}
\startdata
I & 0.630 & 0.007\\
II & 0.311 & 0.005\\
III & 0.610 & 0.013\\
IV & 0.547 & 0.006\\
V & 0.530 & 0.005\\
VI & 0.585 & 0.009\\
VII & 0.623 & 0.005\\
\enddata
\end{deluxetable}
\fi

The average radius of the granules $r_{av}$ is calculated using the same approach taken in \cite{sven09} for solar observations, in this case from a power spectrum of the velocity field on the $z=0$ plane (see Fig. \ref{fig-10} for an example of the spectrum as a function of wavenumber for one snapshot of simulation I), where the value on the x-axis corresponding to the peak of the curve provides the diameter of the granules. From Fig. \ref{fig-10} it is immediately seen that the power spectrum has a broad peak, around $D\simeq2.8~\rm{Mm}$, which gives $r_{av}\simeq1.4~\rm{Mm}$ for this simulation.

\ifnum\apjournal =0
\begin{figure}[b]
\epsscale{1.0}
\plotone{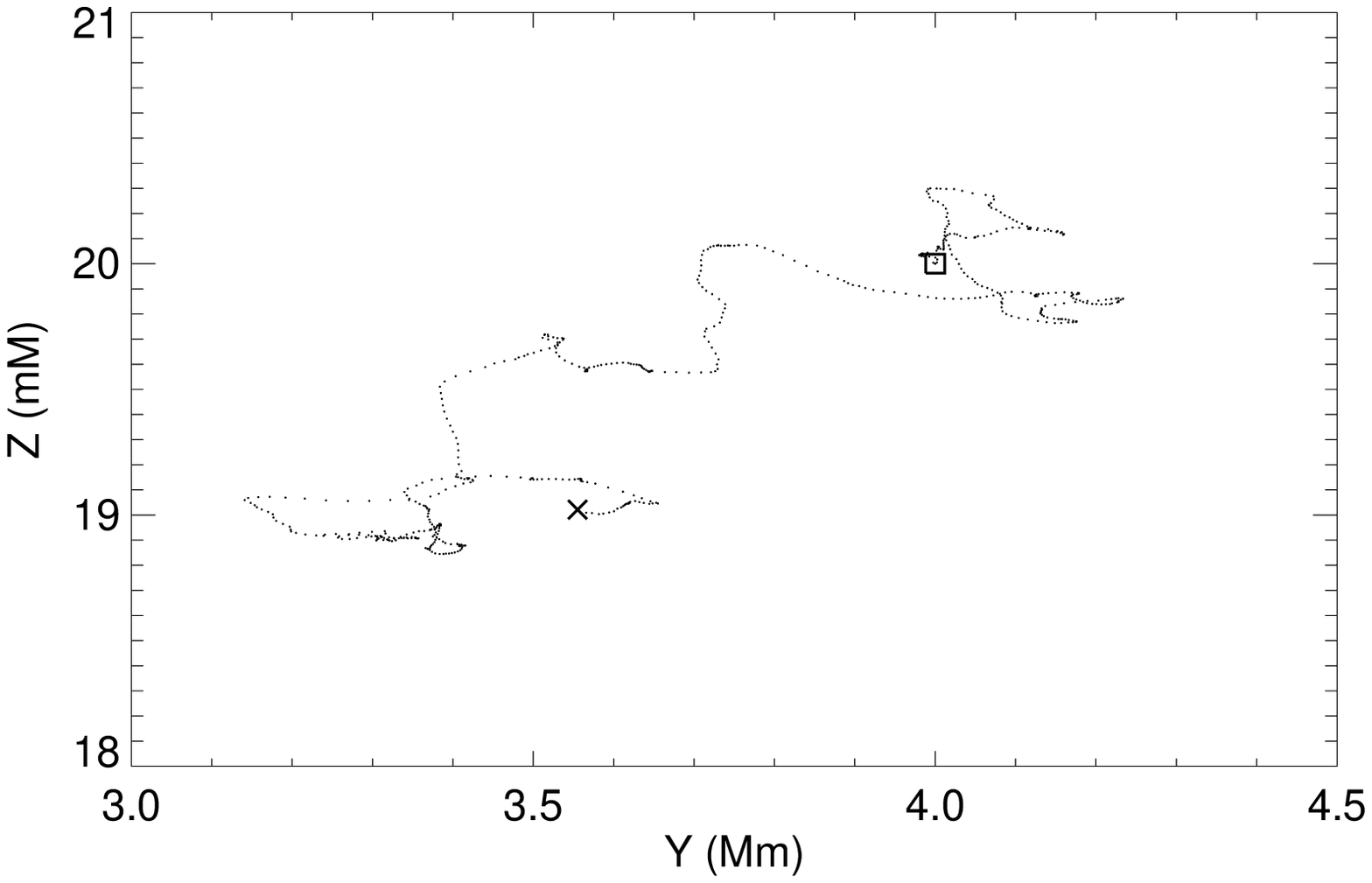}
\caption{Random motion of the footpoint of a magnetic field line, at the bottom of the box. The square symbol marks the initial position and the `X' shows the position of the footpoint at the end of the simulation I.\label{fig-8}}
\end{figure}

\begin{figure}
\epsscale{1.0}
\plotone{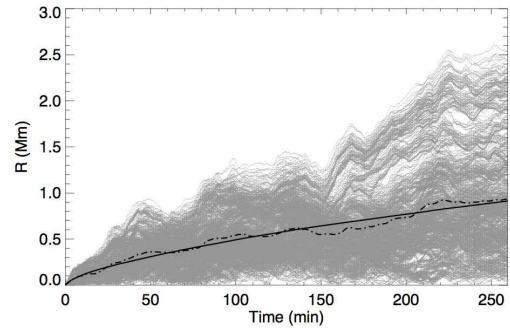}
\caption{Distance $R$ from the original location as a function of time for 1024 footpoints in simulation I. The dash dotted  line corresponds to the average distance and the solid line represents the fit assuming a fitting function $R=Ct^p$.\label{fig-9}}
\end{figure}
\fi

The average velocity $\bar v$ inside the granules is then calculated from the velocity profile (Eq. \ref{eq:velprofile} and Fig. \ref{fig-11}) using the average radius from the powerspectrum of the velocity field:
\begin{equation}
\bar v=\frac{1}{r_{av}}\int_0^{r_{av}} v_h \; dr
\end{equation}

The average  lifetime $\bar \tau$ of the granules is found by looking at the distribution of the number of granules according to their development. Figure \ref{fig-12} shows an example (simulation I) which shows that most of the granules are living within 3.5 minutes around the instant of maximum development ($t-t_{0i}$), which will correspond to an average lifetime of 7 minutes.

Initially $\ell$ is set equal to the product $\bar v \bar \tau$. In case the product  $\bar v \bar \tau$ is greater than $r_{av}$, $r_{av}$ is taken as the step length.

Table~\ref{tab-4} organizes all these quantities discussed above for our simulations.

\ifnum\apjournal =0
\begin{figure}[htp]
\epsscale{1.0}
\plotone{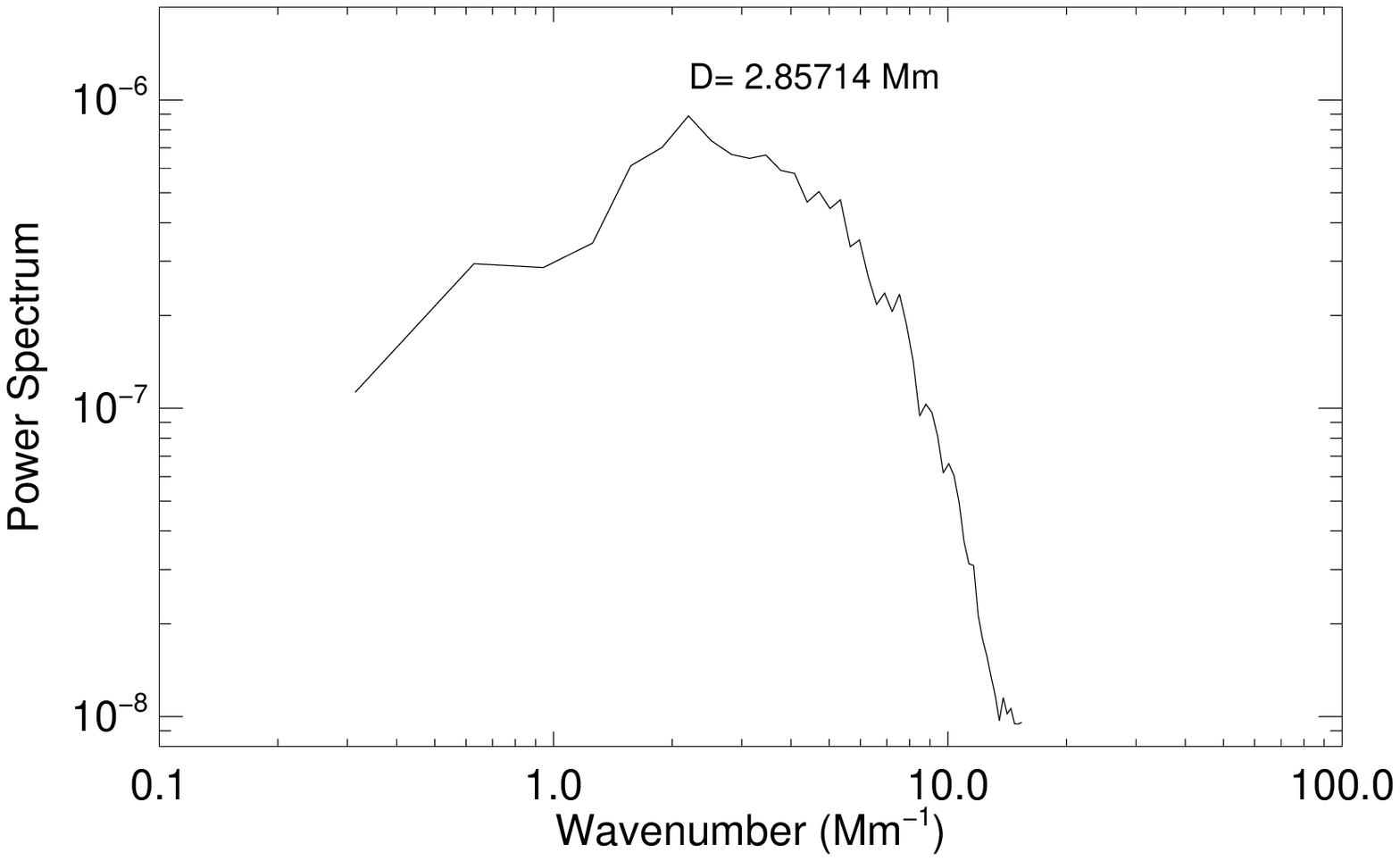}
\caption{Velocity power spectrum vs. wavenumber for a snapshot of simulation I. The figure shows the peak as a sign of the granulation pattern at the bottom of the box.\label{fig-10}}
\end{figure}

\begin{figure}[htp]
\epsscale{1.0}
\plotone{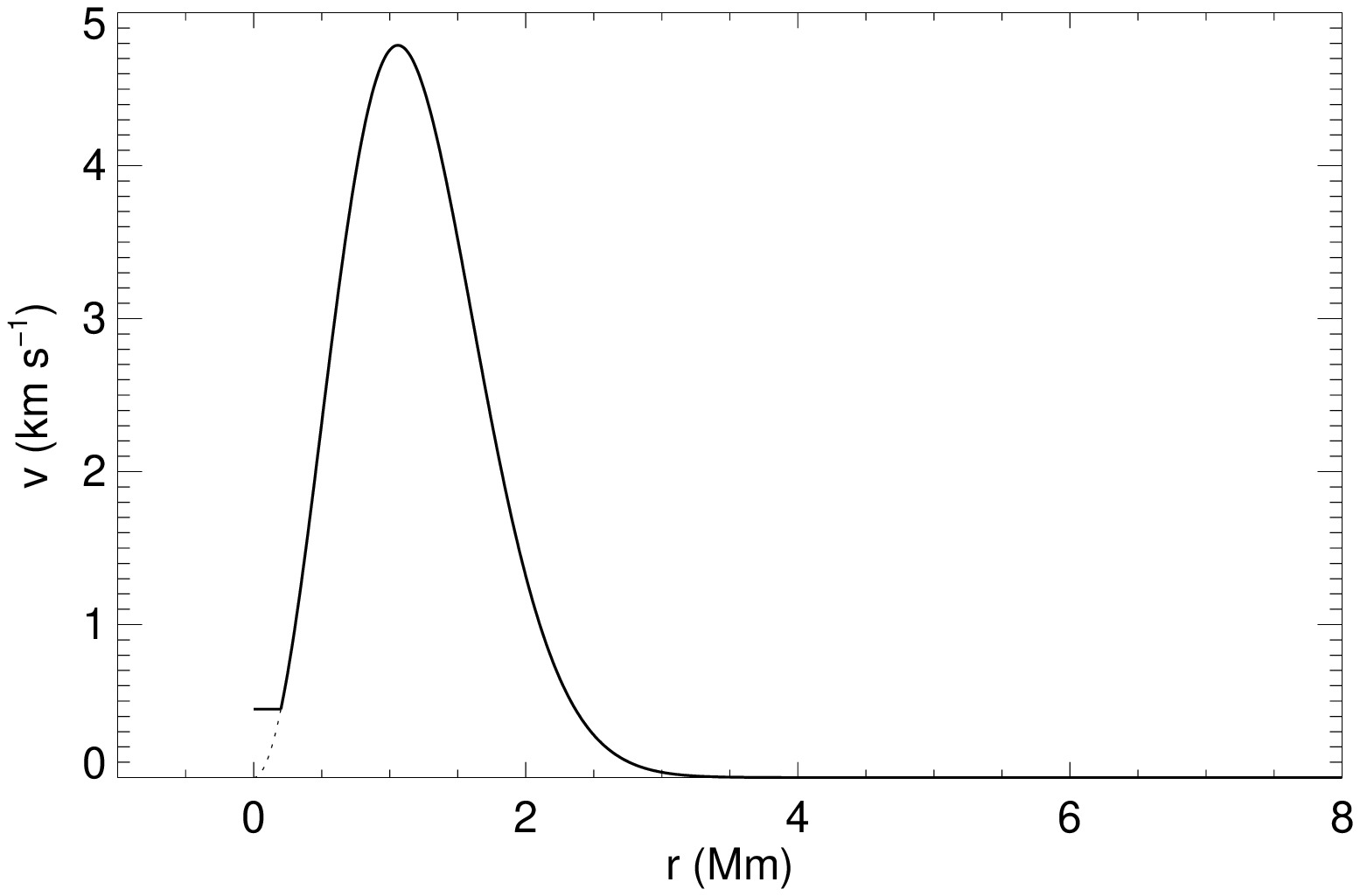}
\caption{Example of the velocity profile of a granule belonging to simulation I. The velocity can never be zero inside the code in order to keep the photospheric motions, so instead of the dashed portion of the plot a constant value of $v=0.2~\rm{km~s^{-1}}$ is assumed.\label{fig-11}}
\end{figure}
\fi

\subsection{Turn over point and critical angle}
Figure \ref{fig-1314} (right) shows the correlation between the granular size and the time of the turn-over point of the simulations. A plausible explanation of this behavior consistent with Parker's model and the random motion of the footpoints relies on the critical angle and on the fact that footpoints must be moved a certain distance from their original location in order to reach that angle. The properties of the random motion means that the total distance traversed by the footpoints is more sensitive to the step length than the number of steps. All simulations of set 1 (where the size of the granules is changed keeping the other initial parameters constant) have approximately the same step length but one has to be careful as the size of granules affects the results. We have been assuming the center of the granule as the original location of footpoints which is not necessarily true of course. Eventually a footpoint may start to be dragged very close to the intergranular lane so it cannot travel the distance equivalent to the step length. In this first set of simulations $\ell\simeq1.10~\rm{Mm}$ which means that for a footpoint to be able to move this distance it has to be located at least one step length from the intergranular lane. If the footpoint starts to move from a point farther away from the intergranular lane, its motion will be limited to $\ell$ by the lifetime of the granule. There will be then an area of the granule around the center where footpoints can start the motion with high probability of traveling along $\ell$. This area inside granules is obviously larger and larger as we increase the size of the granules meaning that, statistically, the real step length will closely match the value found for larger granules and will progressively reduces as we decrease the granular size. 
 Our values of step length correspond in this way to the maximum value the step length can take. Taking the size of the granules into account it is reasonable to think then that our value of the step length will diverge more and more from the real value as we decrease the size of the granules. This can justify the inverse correlation observed between the granular size and the time for the turn-over point: smaller granules will have a smaller and smaller value of step length and then the equilibrium between current generated and released in the corona will happen at a later time when compared with larger granules.
\ifnum\apjournal =0
\begin{figure}[t]
\epsscale{1.0}
\plotone{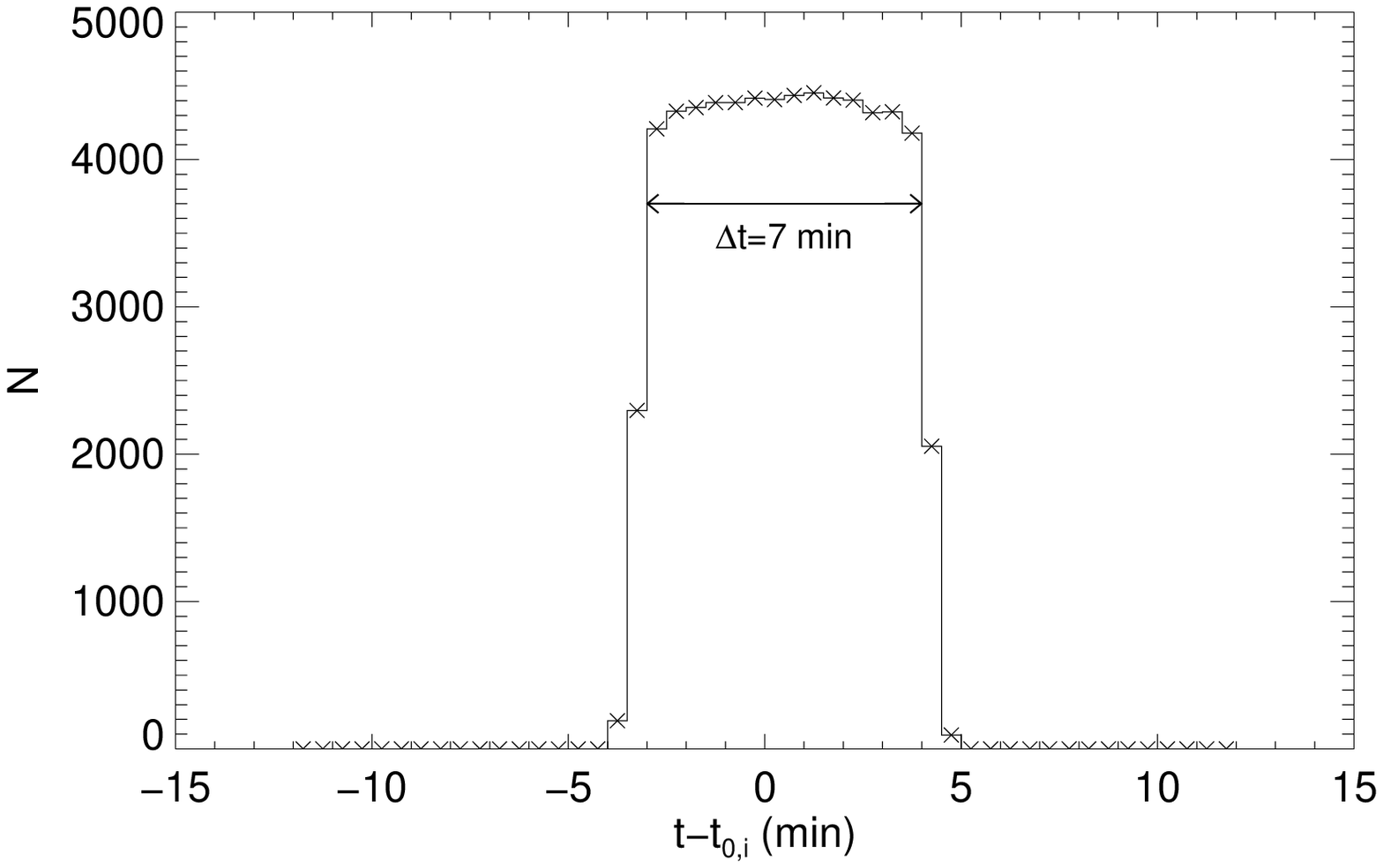}
\caption{Histogram showing the distribution of granules according to their development in simulation I. Negative values on the x-axis correspond to the period pre-full development while the positive values represent post-full development phase.\label{fig-12}}
\end{figure}
\fi

\ifnum\apjournal =0
\begin{deluxetable}{cccccc}[b]
\tablecaption{Average granular size ($r_{av}$), step length ($\ell$), average velocity ($\bar v$), average lifetime ($\bar \tau$) and average number of random steps ($N$) for each experiment \label{tab-4}}
\tablehead{
\colhead{Simulation} & \colhead{$r_{av}$ (Mm)} & \colhead{$\ell$ (Mm)} & \colhead{$\bar v~\rm{(km \cdot s^{-1})}$} & \colhead{$\bar \tau~\rm{(min)}$} & \colhead{N}
}
\startdata
I & 1.26 & 1.10 & 2.63 & 7 & 37.1\\
II & 1.66 & 1.10 & 2.29 & 8 & 31.4\\
III & 2.63 & 1.11 & 2.17 & 8.5 & 15.1\\
IV & 1.32 & 1.32 & 2.58 & 11 & 29.0\\
V & 1.29 & 1.29 & 2.63 & 17 & 24.4\\
VI & 1.25 & 1.25 & 5.13 & 7 & 37.3\\
VII & 1.30 & 1.30 & 8.61 & 7 & 46.7\\
\enddata
\end{deluxetable}
\fi

Figure \ref{fig-1516} (right) shows the variation of the instant of turn-over point as a function of granular lifetime. Here the correlation is linearly increasing. According to the random walk theory there should be no dependence on the granular lifetime. Nevertheless we should not exclude the influence of the adjacent granules and their lifetime. For shorter lifetimes the topology of velocity field changes faster, and the magnetic field lines trapped in intergranular lanes will be moved sooner due to the faster changing velocity field topology. A second effect is that the electric current density is lower as we will see in Sec. \ref{sec:avg-el-current}.

The time of turn-over point is also influenced by the velocity field and the correlation is shown in Figure~\ref{fig-1718} (right). It is obvious that as we increase the parameter $v_{0}$ in the simulations, the average step velocity ($\bar v$) also increases approximately by the same amount, which will induce an increase of the step length. Associated with this fact, the electric current produced also dramatically increases and the time needed to balance this process by dissipating the larger amount of current will be longer than for low velocities.

At this point there are enough elements to give an estimate (in a rough way as we are doing some strong assumptions) for the critical angle introduced by Parker in his model. This critical angle should correspond to the average deviation from the vertical of the magnetic footpoints when the dissipation starts to cancel the transverse component of the magnetic field at the same rate it is created by the drag of footpoints in the photosphere. In this study that moment will coincide approximately with the time of turn-over point, when after the fast initial growth of electric current density we observe a stabilization of this quantity caused by the balance between creation and dissipation of electric current.

Using the values of the time of turn-over point, time step and step length and applying them on Eq. \ref{eq:randomtheory}, one can calculate the average displacement of the magnetic footpoints from its original location
\begin{equation}
\label{eq:displ_top}
R_{turn}  = \ell \sqrt {\frac{{t_{turn} }}{\bar \tau }}
\end{equation}
where $t_{turn}$ is the time of turn-over point and $\bar \tau$ the time step, so $t_{turn}/\bar \tau=N_{turn}$.

The estimated deviation angle from the vertical will then be 
\begin{equation}
\label{eq:dev_angle}
\theta _{est}  = \arctan \left( {\frac{{R_{turn} }}{H}} \right)
\end{equation}
with $H$ being the height of the box, $H=20~\rm{Mm}$. Results are indicated on Table~\ref{tab-5}.

Due to all the assumptions taken along this work, it is hard to find correlations between the angle and the initial parameters, or at least to put a high level of trust in them. Nevertheless one can observe that there are no huge variations of the estimated angle which varies from 5.5 to approximately 12.5, thus leading to an average value around 10 degrees. This is not very far from the 14$^{\circ}$ postulated by Parker and could mean that our assumptions are justifiable. There is a particular discrepancy in set 1 of experiments where the estimated angle decreases from 8.76$^{\circ}$ in simulation I to 5.50$^{\circ}$ in simulation III. According to \ref{eq:dev_angle}, the critical angle is proportional to the displacement at the moment of turn-over point, which is proportional to the step length from \ref{eq:displ_top}. We should expect a smaller angle for smaller granular size as the turn-over point is reached faster leading to a shorter displacement and the calculated step length is about the same in all these 3 simulations. However it was already explained at the beginning of this section that the true step length should be smaller for smaller granules, so this fact would uniform the critical angle, making it approximately the same for the three simulations.

\subsection{Average electric current in the atmosphere}\label{sec:avg-el-current}
The energy dissipated by currents in the atmosphere originates as Poynting flux injected into the magnetic field by the photospheric motions. The amount of dissipated energy in the atmosphere must therefore be proportional with the injected energy flux when a statistical equilibrium is reached. If this is correct there should exist simple relations between the granular parameters and the average electric current in the atmosphere. \citet{parker83a} estimated the Poynting flux injected into the atmosphere:
\begin{equation}
P = \frac{v B_{\perp} B}{4\pi}=\frac{B^2v^2 \Delta t}{4\pi H} \label{eq:poynting}
\end{equation}
where $v$ is the velocity which the magnetic footpoint is moved with (measured in km$/$s) during the time interval  $\Delta t$, $B$ is magnetic field strength, $B_{\perp}$ is the field strength of the transverse component of the magnetic field generated by the movement of the footpoint and $H$ is the height of box. As there is only a dependency on the flow velocity, the relations should here be rather simple. It is though important to notice that the velocity here is the effective velocity, meaning the velocity that carries a magnetic footpoint further away from it initial position in a straight line. It is therefore not that simple to estimate the effect of the variables defined earlier,  $r_{av}$, $\bar \tau$, and $\bar v$.

Figure \ref{fig-1314} (left) shows the dependence of the average current density in the atmosphere as a function of the average granular size and the good linear fit between these two quantities: the current density generated goes up as the average size of the granules increases. The step length calculated for set 1 is roughly the same but because of the reason already pointed out in the previous discussion about the variation of the instant of turn-over point with granular size, one should expect a smaller value of the step length than the calculated value for smaller granules. So in a photosphere with larger granules the motions of the plasma will cause a higher displacement in a straight line of the footpoints of the magnetic field lines, twisting the magnetic field more than in the case of smaller granules and thus producing more electric current. With the module of the Poynting vector proportional to the transverse component of the magnetic field, $B_{\perp}$, a larger granule,  produces higher $B_{\perp}$ by creating a stronger twist of the magnetic field and therefore produces an increase of the energy flux. 

Figure~\ref{fig-1516} (left) evinces a correlation between the average granular lifetime and $J_A^2$.
Magnetic footpoints caught inside a granule that lives longer follow a straight path radially inside the granule for a longer time, and therefore, the effective velocity is higher. This happens as long as the granules are large enough for the magnetic footpoints not to reach the edges of the granules within the 
granular lifetime, but in this set 2 of simulations the granular size is kept constant and then the influence of the size of the granules is the same for the three experiments of this set. On the other hand, inside short-lived granules footpoints will move straight in shorter time steps decreasing the effective velocity and consequently the $J_A^2$.

\ifnum\apjournal =0
\begin{deluxetable}{cccccc}[b]
\tablecaption{Values of the average displacement at the turn-over point ($R_{turn}$) and estimates for the critical angle($\theta_{est}$) \label{tab-5}}
\tablehead{
\colhead{Simulation} & \colhead{$R_{turn}$ (Mm)} & \colhead{$\theta_{est}$}
}
\startdata
I & 3.08 & 8.76$^\circ$\\
II & 2.35 & 6.71$^\circ$\\
III & 1.92 & 5.50$^\circ$\\
IV & 3.89 & 11.0$^\circ$\\
V & 3.83 & 10.9$^\circ$\\
VI & 3.71 & 10.5$^\circ$\\
VII & 4.44 & 12.5$^\circ$\\
\enddata
\end{deluxetable}
\fi

In Figure~\ref{fig-1718} (left) we show an identical scenario for the variation of the $J_A^2$ as functions of the average velocity field at the photosphere. 
From equation \ref{eq:poynting} it seems that there should be a correlation between the total current squared and the velocity squared, but again the velocity in equation \ref{eq:poynting} is the straight line velocity while the velocity experienced by the magnetic footpoints are not the effective straight line velocity. The correlation between the $J_A^2$ and the velocity can then be explained by the random walk nature of the velocity field. 

One can think, based on expression \ref{eq:theta} that a similar correlation should be obtained between the average current density or the turn-over point and the product $\bar v \bar \tau$. Following the previous discussed relation of average current density and the angle $\theta$, the latter should increase with $\bar v \bar \tau$, and the same should apply for the turn-over point. Figure~\ref{fig-1920} shows this behavior, although it is not possible to linearly fit all the data points. This non-linear increase of both the average current density and turn-over point with $\bar v \bar \tau$ implies that these variables are not connected in a simple way, and that the relation between the amount of electric current arising in the corona and the deviation angle $\theta(t)$ to the vertical should be more complex. This is probably again an effect of the difference between the $\bar v$ and the velocity used in equation \ref{eq:poynting}.

\ifnum\apjournal =0
\begin{figure*}
\epsscale{1.0}
\plottwo{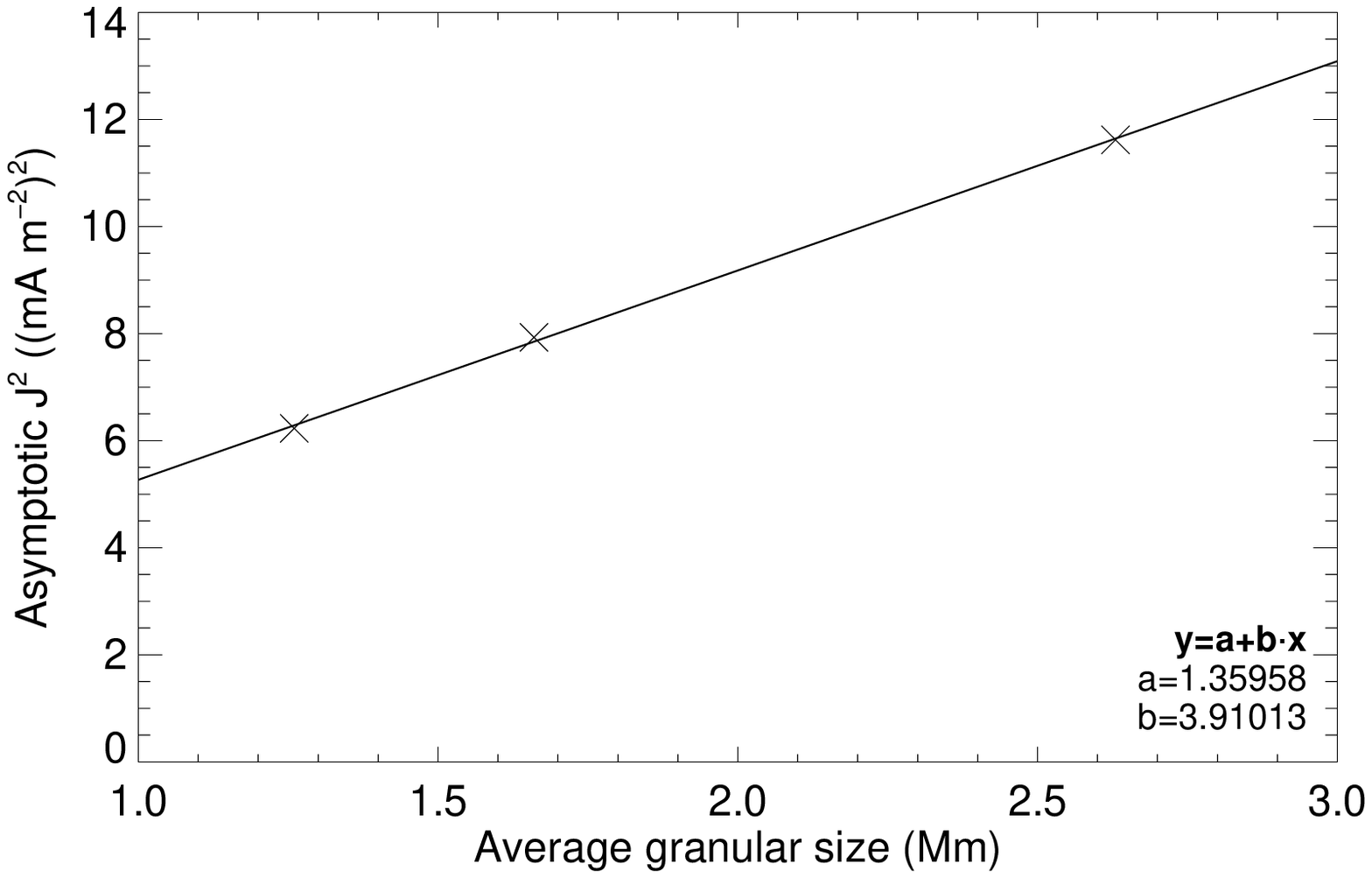}{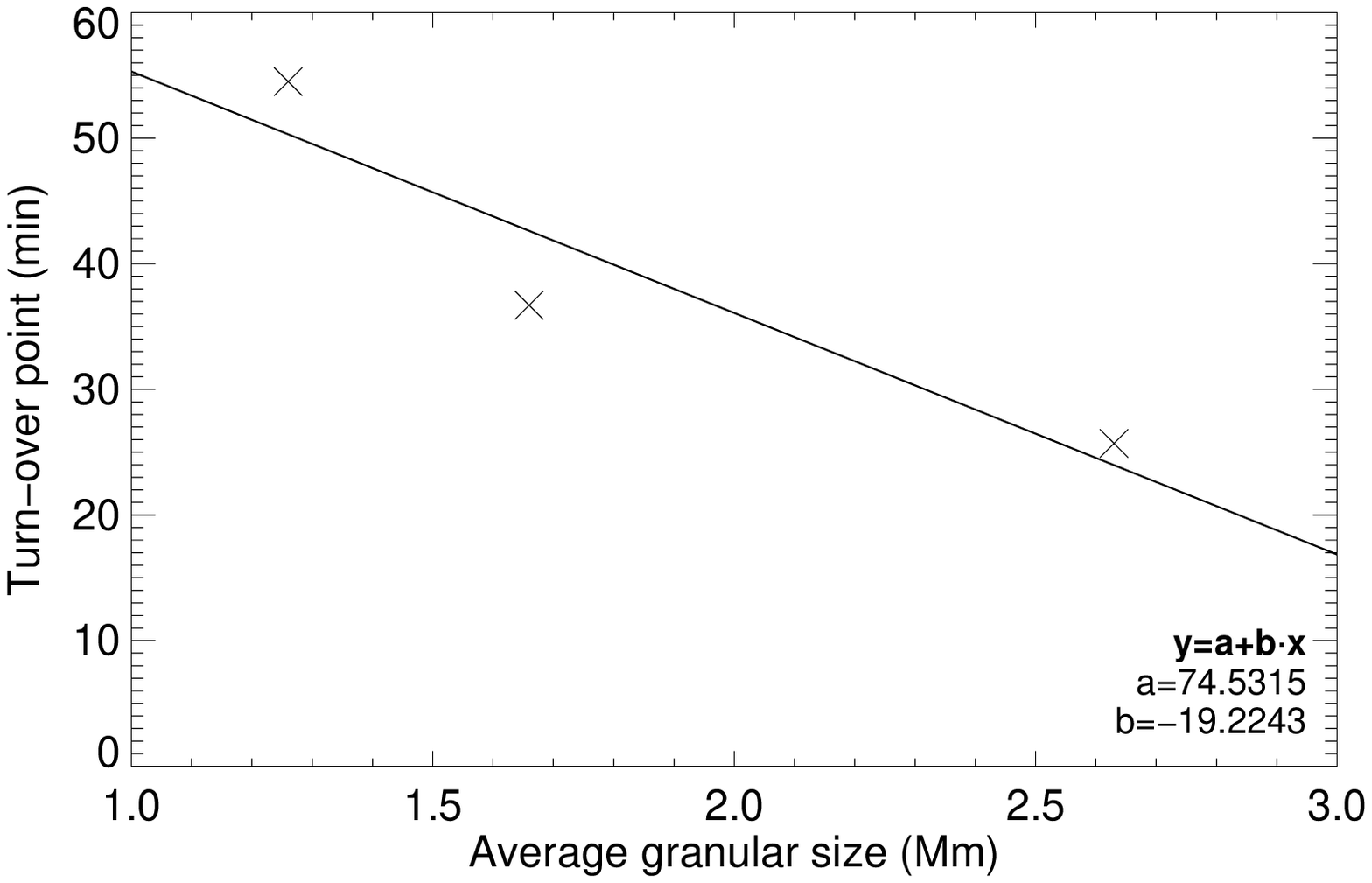}
\caption{$J_A^2$ (left) and turn-over points (right) as a function of the average granular size, from simulations of set 1.\label{fig-1314}}
\end{figure*}

\begin{figure*}
\epsscale{1.0}
\plottwo{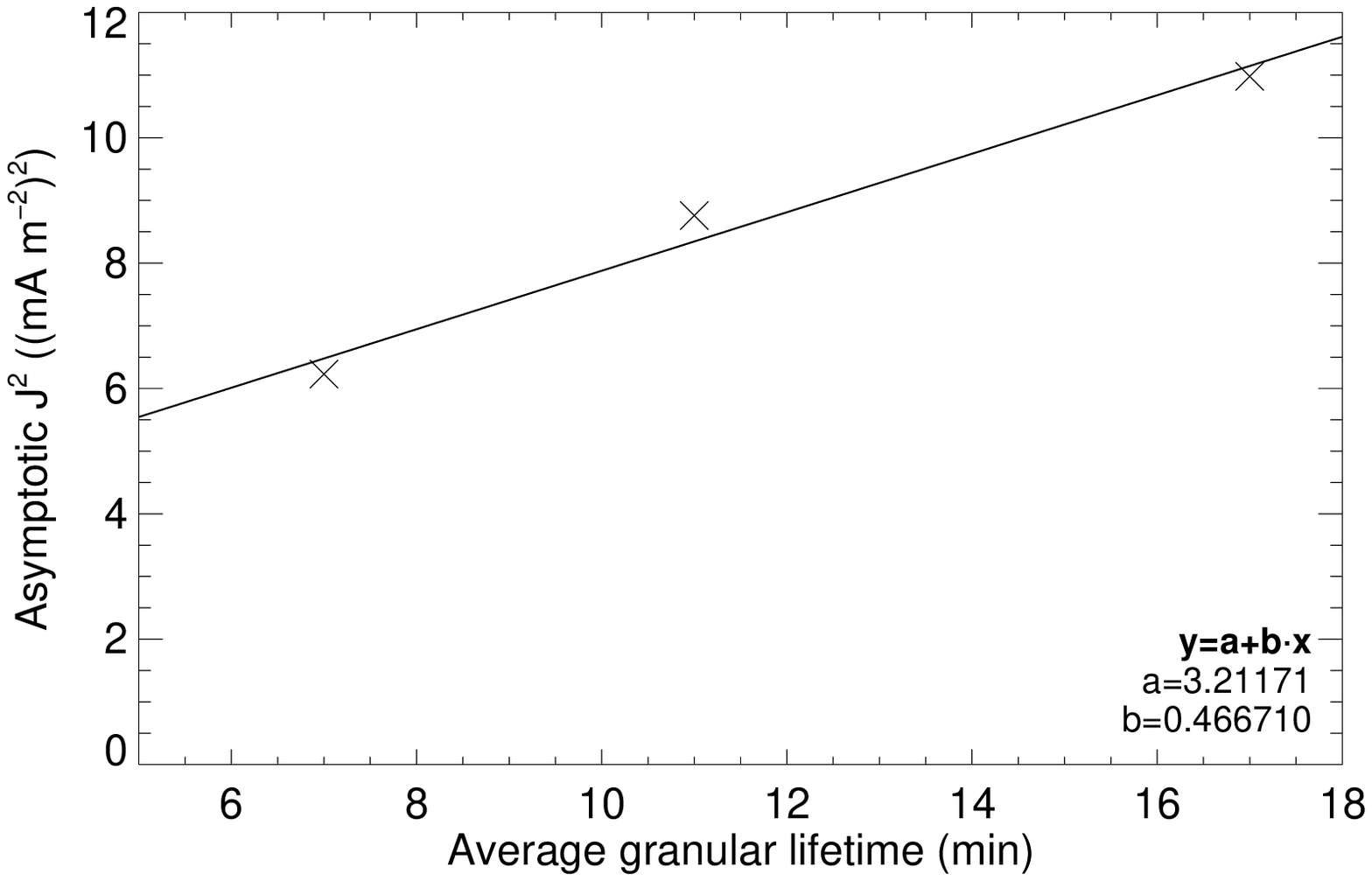}{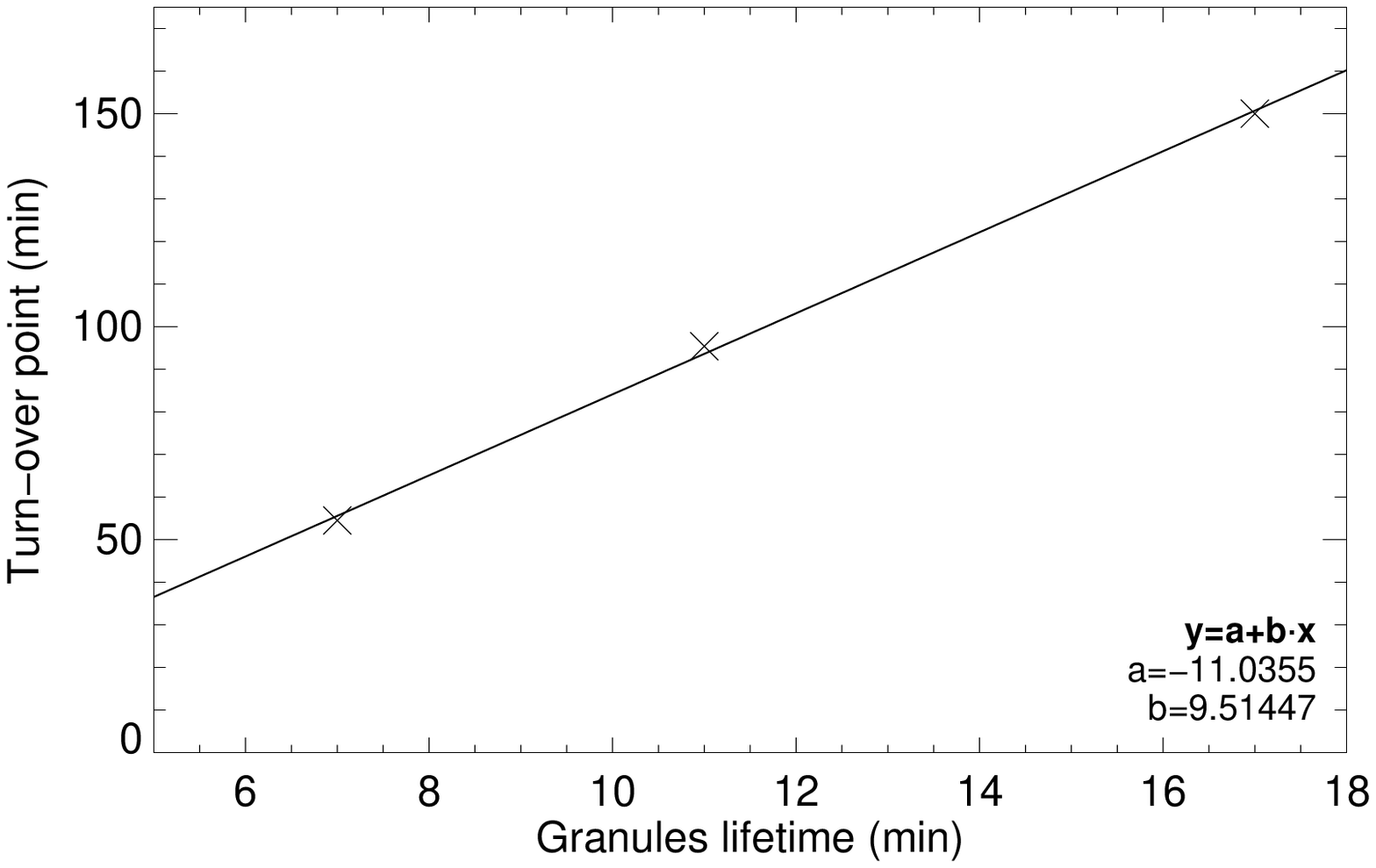}
\caption{$J_A^2$ (left) and turn-over points (right) as a function of the average granular lifetime, from simulations of set 2.\label{fig-1516}}
\end{figure*}

\begin{figure*}
\epsscale{1.0}
\plottwo{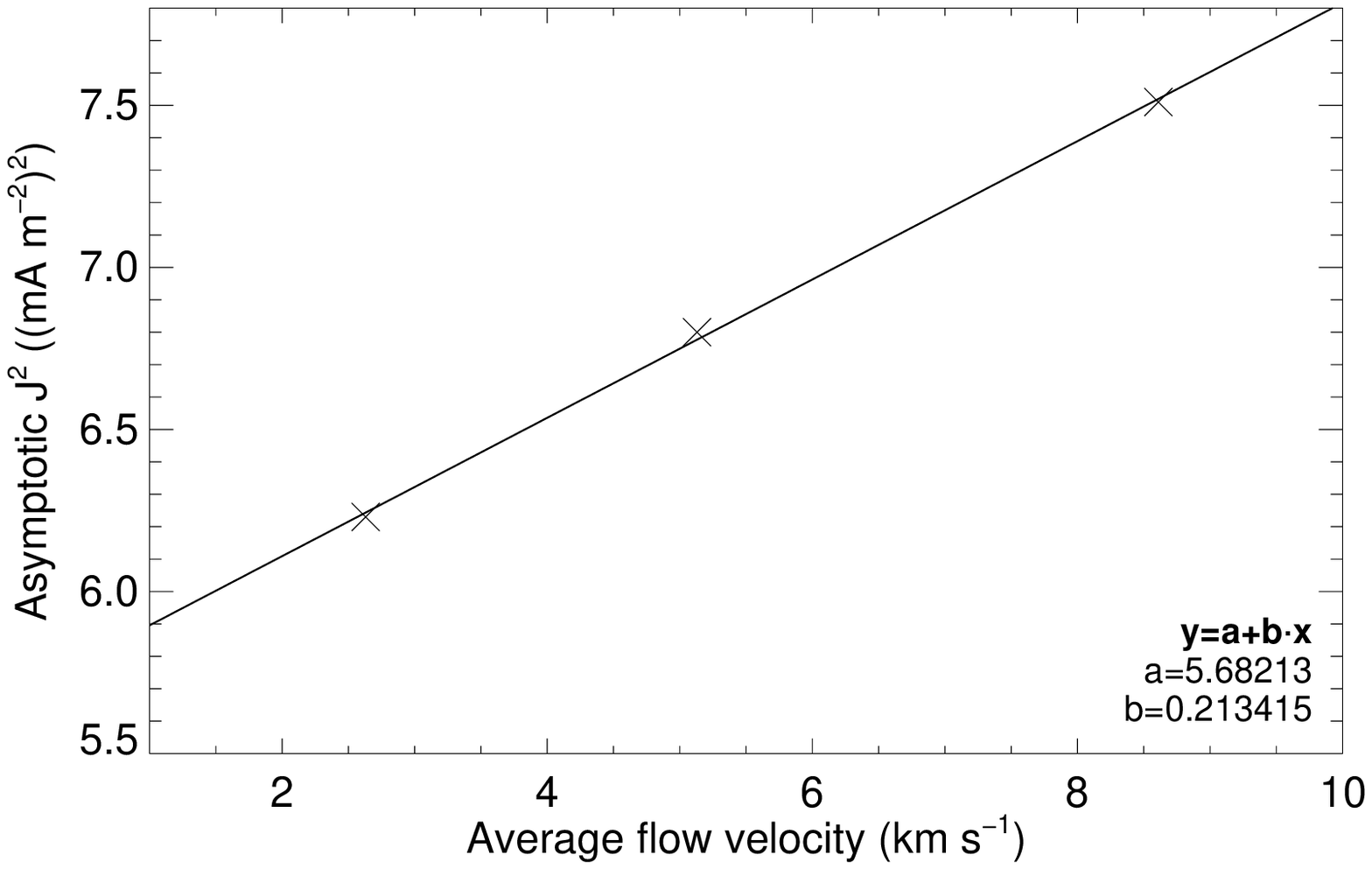}{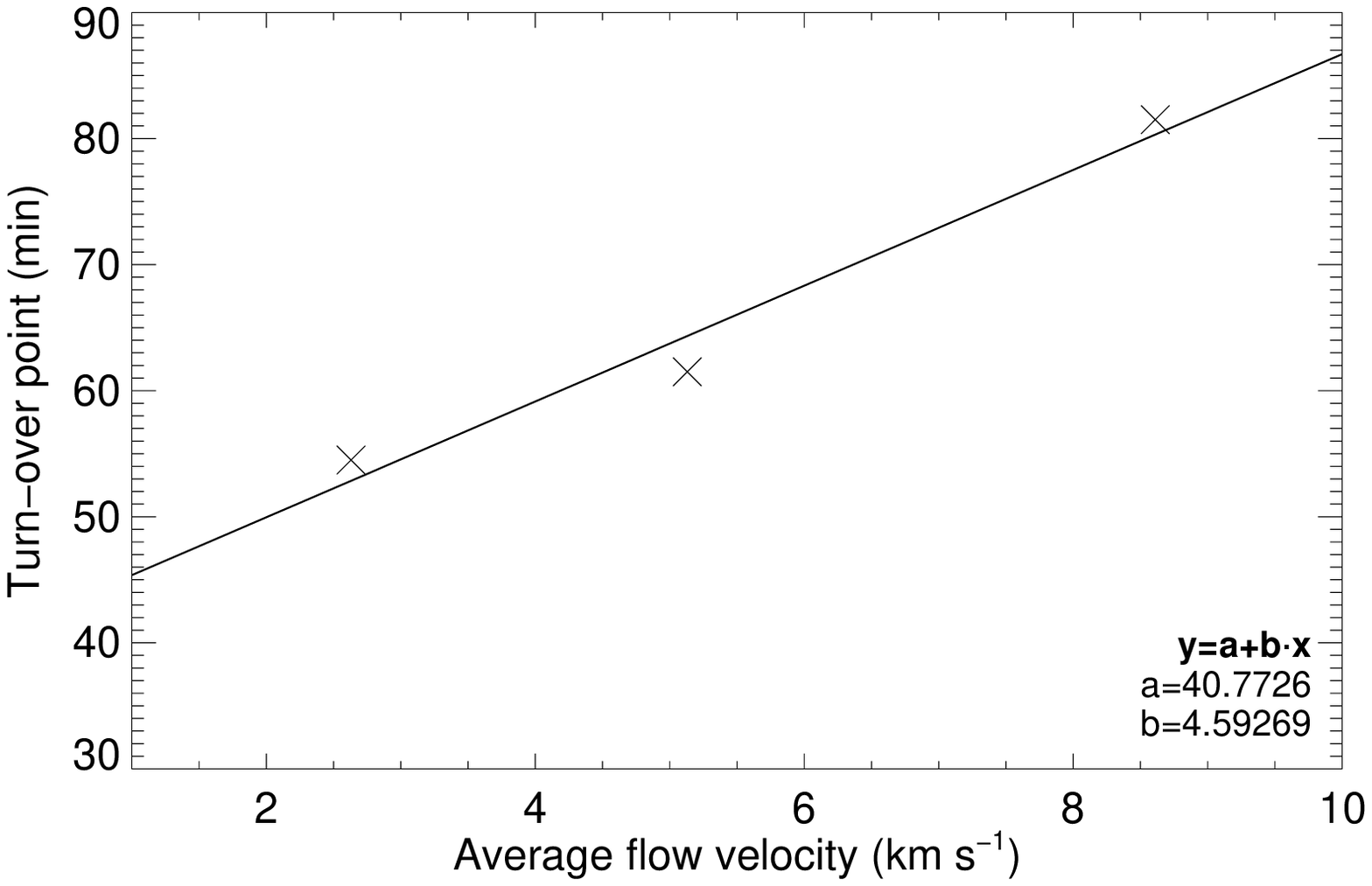}
\caption{$J_A^2$ (left) and turn-over points (right) as a function of the average flow velocity, from simulations of set 3.\label{fig-1718}}
\end{figure*}

\begin{figure*}
\epsscale{1.0}
\plottwo{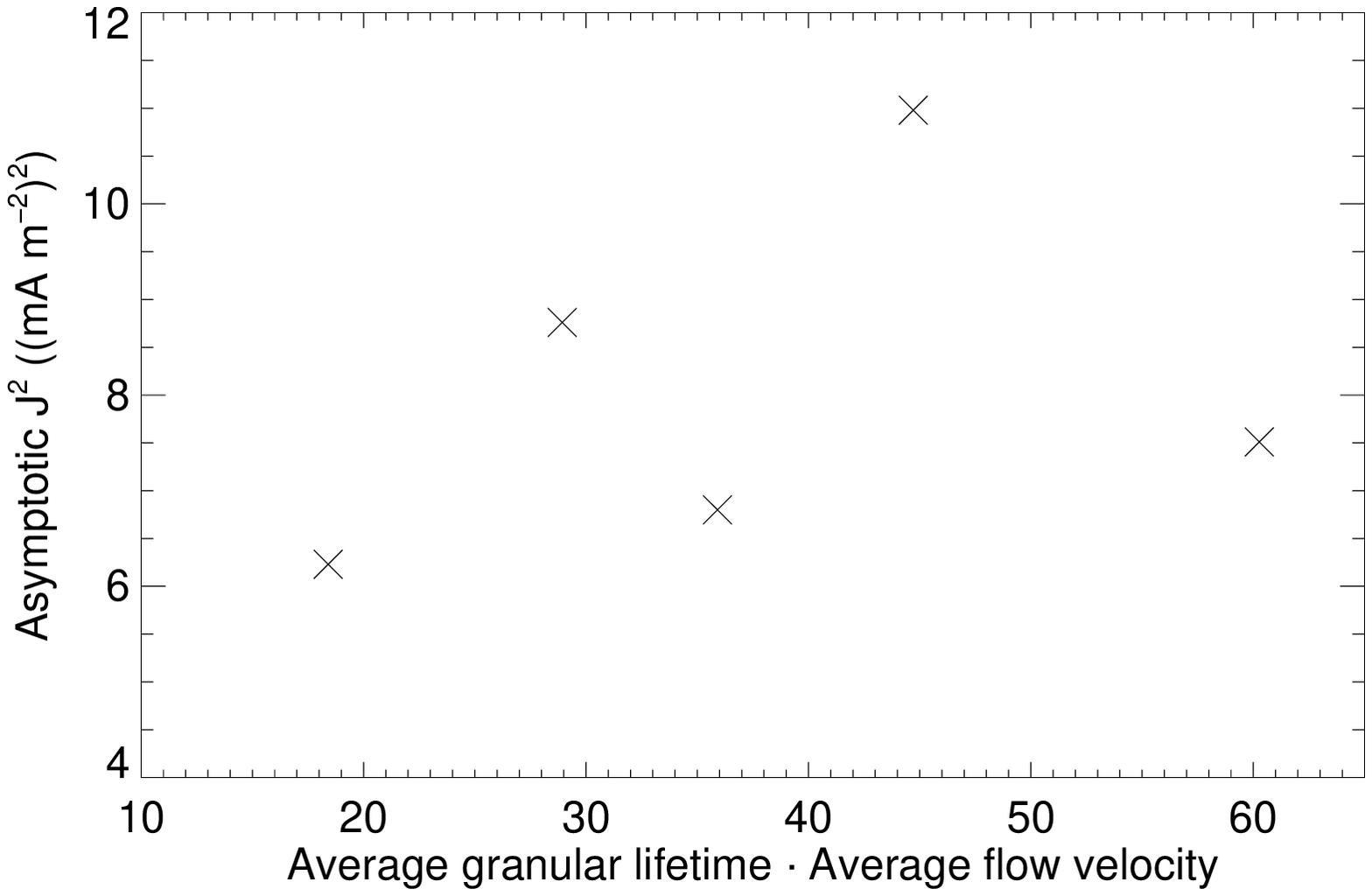}{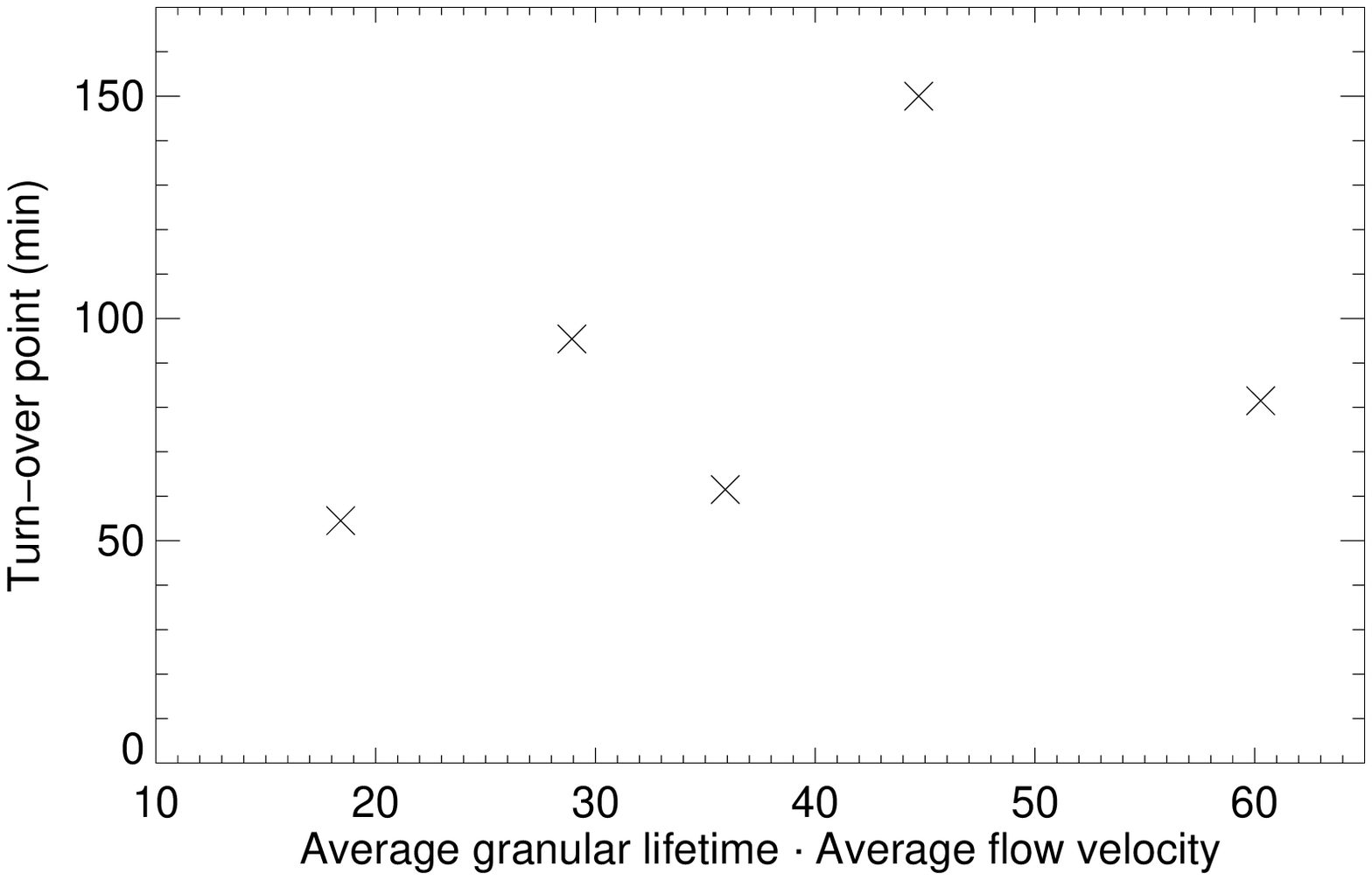}
\caption{$J_A^2$ (left) and turn-over points (right) as a function of the average granular lifetime $\times$ average flow velocity, from simulations I, IV, V, VI and VII.\label{fig-1920}}
\end{figure*}
\fi

\section{Conclusion}

At the present time, we have the opportunity to deploy significant computing power available to explore in a deeper way the heating process of stellar atmospheres. Because observations from stars with the same resolution we get for the Sun are impossible to obtain due to the distance, numerical simulations stands as the only way to explain some of the more complex behaviours observed on stars.

In this work we perform a number of simulations with the intention of studying how photospheric parameters affect the amount of electric current density in the corona. In particular we analyze three parameters, the size of granules, lifetime and the velocity of the plasma at the surface, using a $20^3~\rm{Mm}^3$ box with a stellar quiet photosphere at the bottom. We have seen that tuning these three variables at the surface and thus, simulating different environments at the stellar photosphere, we get an initial sharp increase of the electric current density induced by the motions of the plasma dragging, twisting and shuffling the magnetic field lines. Later, the production of electric current decelerates, reaching a state with average constant value after the turn-over point. Oscillations around the $J_A^2$ are a result of the random behavior of the advection of the footpoints of the magnetic field.

According to our simulations, the time when the electric current density gets to the steady-state is a consequence of the balance between creation and destruction of $B_{\perp}$ as well as the value of the $J_A^2$. Both of them changes as we change the parameters $r$, $\tau$ and $v$. 
Based on our experiments, we conclude that an increase of any one of these parameters will lead to a higher value of the $J_A^2$, and that this relation of direct proportionality can be described with a linear function. Higher values of electric current density at the steady-state are obtained for larger granules as a result of creation of $B_{\perp}$ because the linear radial velocity inside the granules can work longer due to the smaller risk of the magnetic footpoint hitting an intergranular lane. 
The step length, the distance travelled by a magnetic footpoint every random step, plays an important role here as the production and release of current in the atmosphere is closely associated with it. The lifetime of granules is also a factor which affect the electric current density on the corona. Following Parker's idea, the deviation angle $\theta$ is proportional to the time travelled by a footpoint along the same direction. Thus, longer lifetimes mean higher horizontal displacements of the magnetic footpoints which cause an increase of the asymptotic electric current density. The turn-over point is also affected by the lifetime of the granular cells at the photosphere. With short-lived granules the surface of the star changes quickly, the footpoints of the magnetic field are shuffled, twisted and braided faster culminating in earlier turn-over points. 

Parker's argumentation can again be used to explain results of the variation of flow velocity. We have obtained an increase of electric current density at the steady-state as the velocity goes up. As of \ref{eq:transverseB}, the transverse component of the magnetic field is correlated with the flow velocity, so higher flow velocities generate more electric current. 
The resulting Poynting flux is proportional to the driving velocity and at the steady state, that Poynting flux must lead to a total dissipation that is identical to the Poynting flux injected into the magnetic field. 

By doing these experiments, and varying only three parameters of the photosphere, we have shown that these parameters can affect the amount of electric current density in the corona and consequently the energy dissipated. The correlations obtained show that the size of the granules on the stellar surface, their lifetimes and the magnitude of the velocity of the plasma in the granules affect the amount of current in the corona, which have influence on the process of coronal heating. In the context of electric current heating fed by the photosphere these parameters seem to assume particular importance by determining the way the magnetic field is tangled above the photosphere and how energy will be released. Our simulations allow us to make an estimate of the so-called critical angle, first introduced by Parker, to be around 10$^{\circ}$. Although some strong assumptions are made in this work, this value does not look very distant from the 14$^{\circ}$ suggested by \cite{parker88}. 

However, we should also point out some limitations of our model. Although this model is consistent with the ideas supporting the nanoflare coronal heating hypothesis, we cannot resolve single events. \citet{galsgaard96} shows and argues that the resolution is not the important factor as long as it is reasonable, since the Poynting flux is the provider of the energy, and as long as the injection mechanism is resolved, in this case the granules, the dissipation locations will be correct, the total dissipated energy will be correct, but the fine scale structure of the dissipation will not be resolved. 
The Voronoi tessellation splits the bottom of the box into mosaics representing the granules which are governed by certain parameters. This is a simplified picture of what is a real surface of a star. The model used for stratification in the atmosphere is also simple and some details and contributions for the coronal heating could have been missed as a consequence of that.
In the interpretation of the results the way the step velocities and step times are determined (averages in time and space) may lead to uncertain values, resulting then in no definite step length as well. 
In spite of the problems in evaluating the variables we use for the interpretation of the data, the relatively easy explanations of the results, make the values seem viable. 

Some studies have been done aiming to extract empirical relations from observations of cool stars between quantities such as X-ray flux density or magnetic flux density and some parameters associated with stellar properties like the Rossby number or color index. In particular \citet{noyes84} suggested some relations between stellar activity level and a parameter similar to the Rossby number. They associated the Ca II H+K emission with a funtion $g(B-V)$ which approximates the convective turn-over time ($\hat t_C$). Also \cite{vaughan80} showed a correlation between Ca II H+K index and the color index, demonstrating a wide spectrum of activity level on solar-like stars. On other hand, a power law relationship between soft X-ray and the Ca II H+K fluxes, independent of the stellar color, was found by \cite{schrijver92}. Flux in the X-ray band of the spectrum are mainly originated from the coronae of the stars, as a result of coronal heating mechanisms. By taking this studies and relations and combining those with Parker's model for coronal heating which relies on the motions at the surface of the star to originate the energy release on the top of the atmosphere, we believe that stellar properties and ultimately photospheric parameters have some role to play in all this process of coronal energy losses.

As the results show, distinct photospheric parameters result in distinct turn-over points and amount of electric current density on the corona. This work shows that the characteristics of coronae can be influenced by the surface of stars which is directly connected with the magnetic field above the photosphere.

\acknowledgments

This research project has been supported by a Marie Curie Early Stage Research Training Fellowship of the European Community's Sixth Framework Programme under contract number MEST-CT-2005-020395: The USO-SP International School for Solar Physics. This research was also supported by the Research Council of Norway through grant 170935/V30 and through grants of computing time from the Programme for Supercomputing.

We would like to thank the referee for his helpful comments and suggestions to improve the manuscript.


\clearpage

\ifnum\apjournal =1
\begin{figure}
\epsscale{1.0}
\plotone{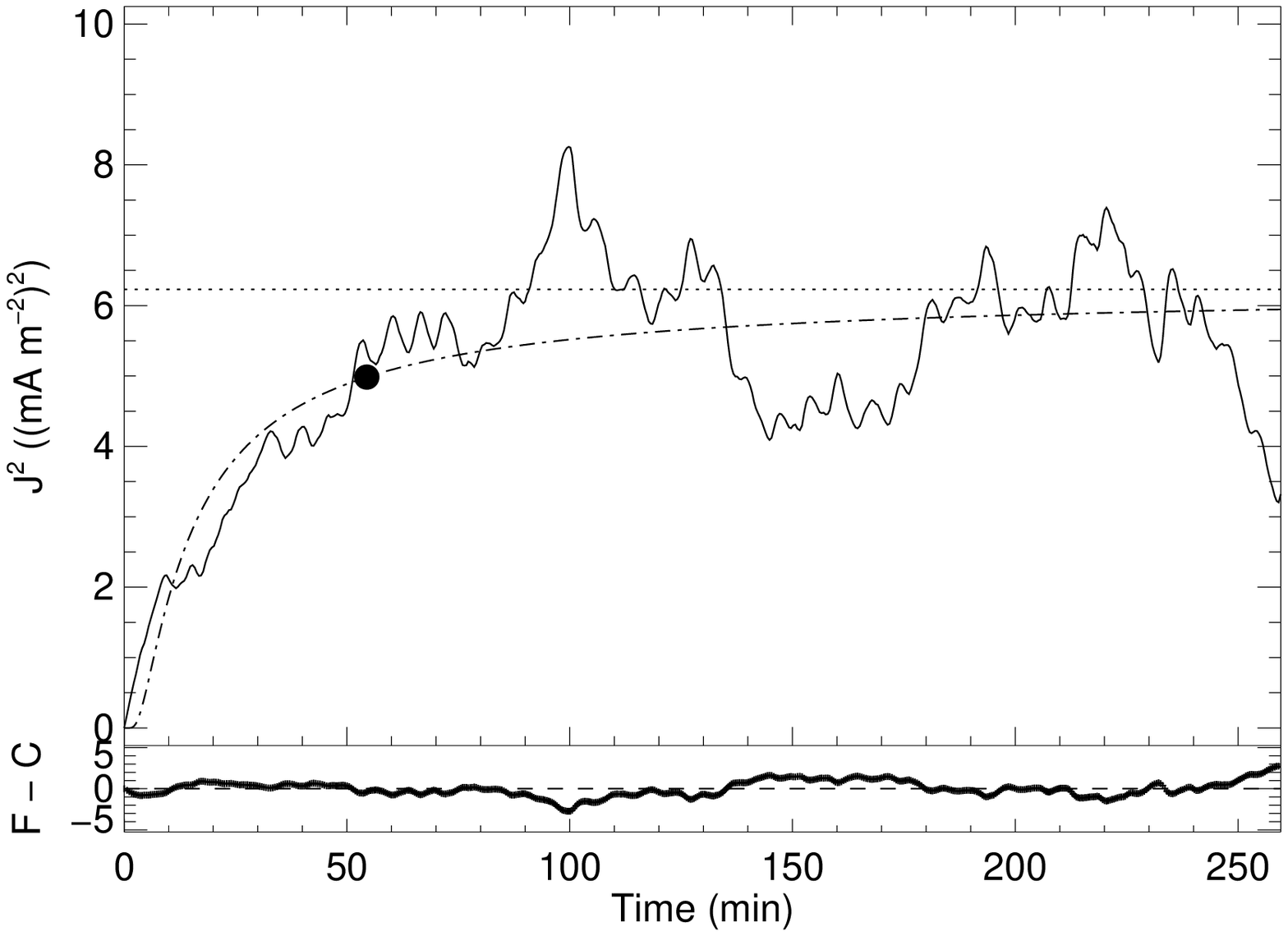}
\caption{Evolution of the average current density in the box as a function of time (solid line) and its curve fitting function (dash dotted line) in simulation I with granules size/ granules lifetime/ flow velocity = 2 Mm/ 5 min/ 3 $\rm{km~s^{-1}}$. Also shown are the $J_A^2$ (horizontal dotted line) and the turn-over point marked with the small filled circle over the fitting curve. The bottom part of the plot shows the residuals in the sense of (F-C: Fitted-Computed).\label{fig-1}}
\end{figure}
\clearpage

\begin{figure}
\epsscale{1.0}
\plottwo{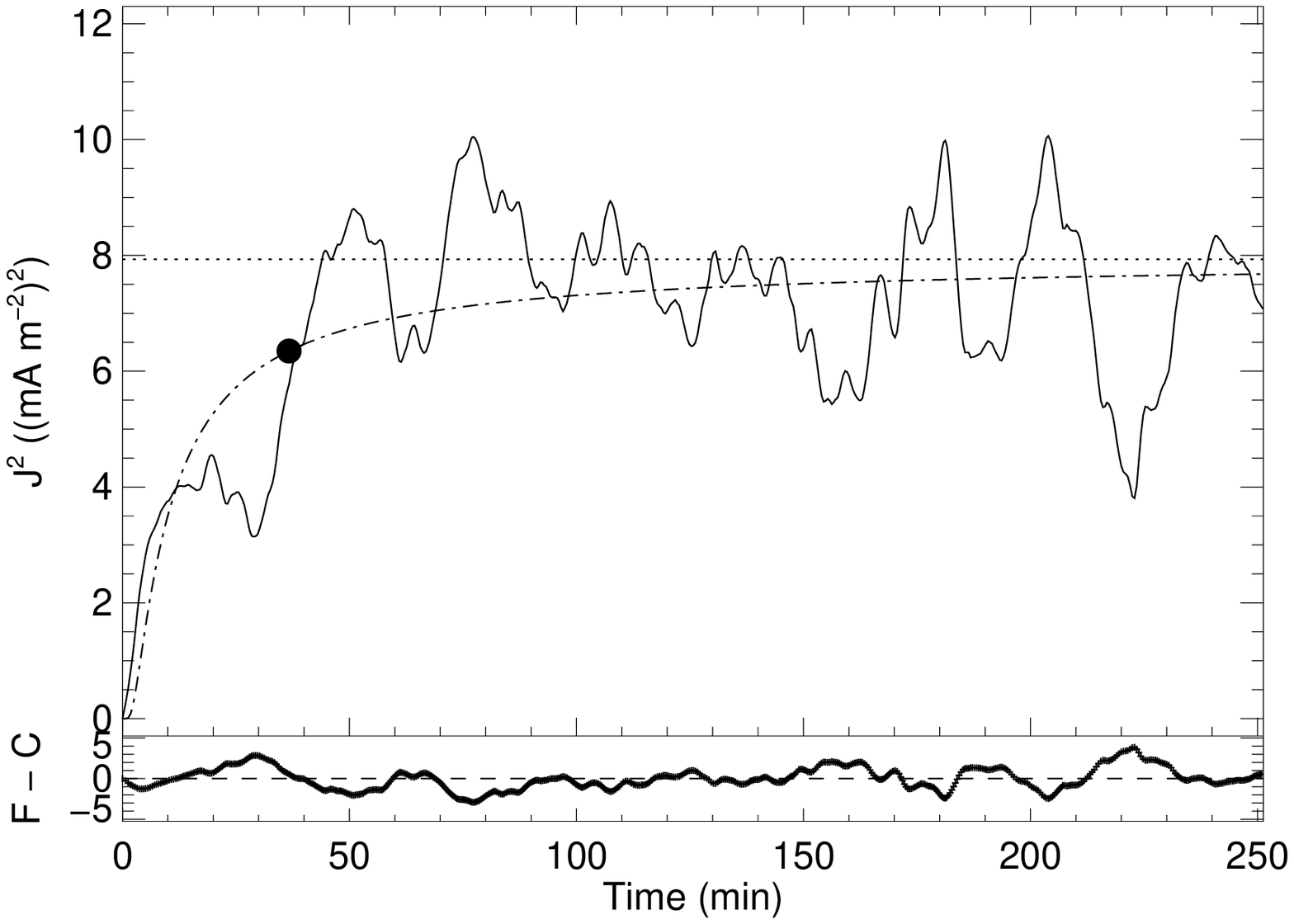}{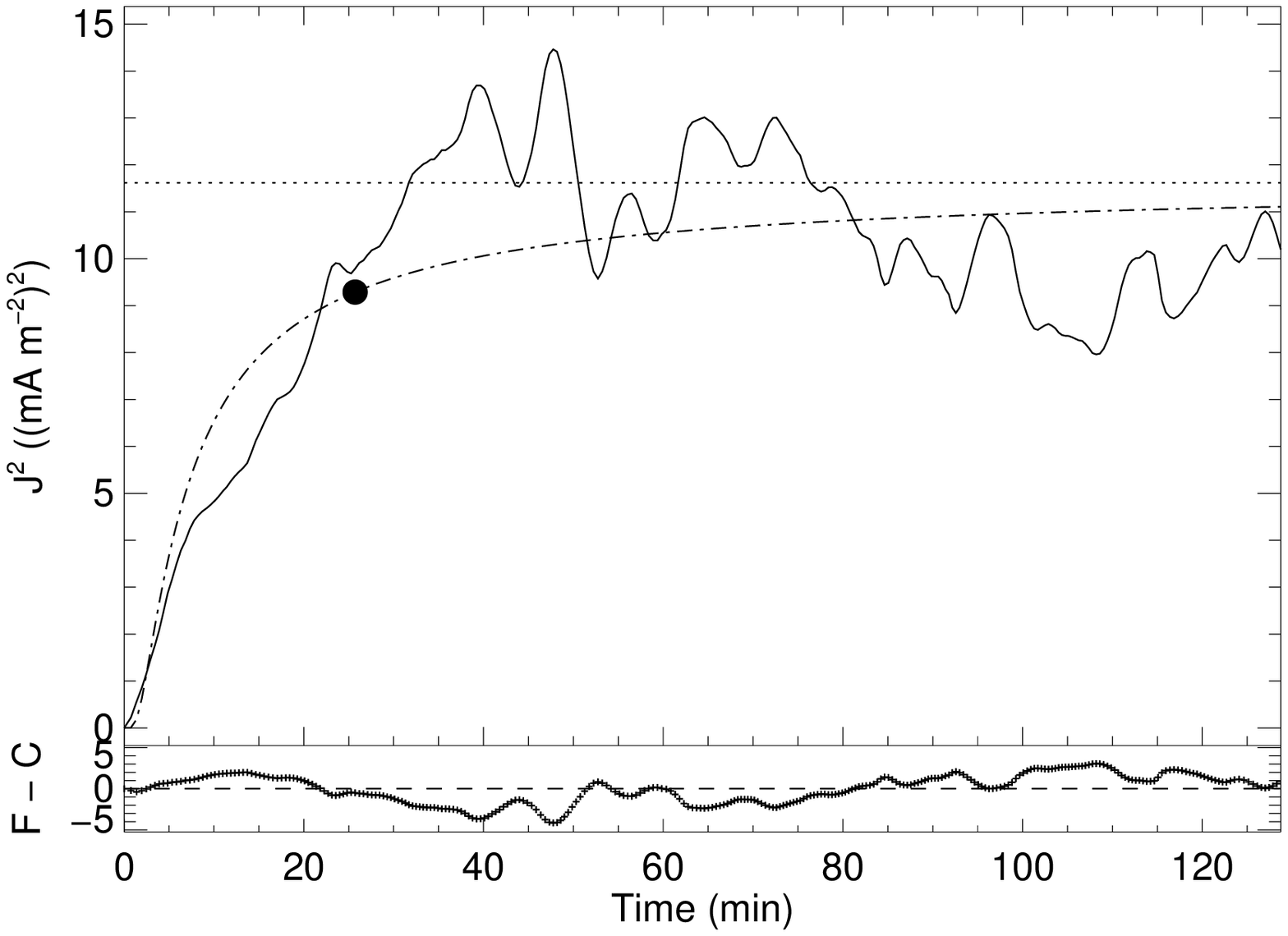}
\caption{Evolution of the average current density in the box as a function of time (solid line) and its curve fitting function (dash dotted line), in simulations II (left) and III (right).\label{fig-23}}
\end{figure}
\clearpage

\begin{figure}
\epsscale{1.0}
\plottwo{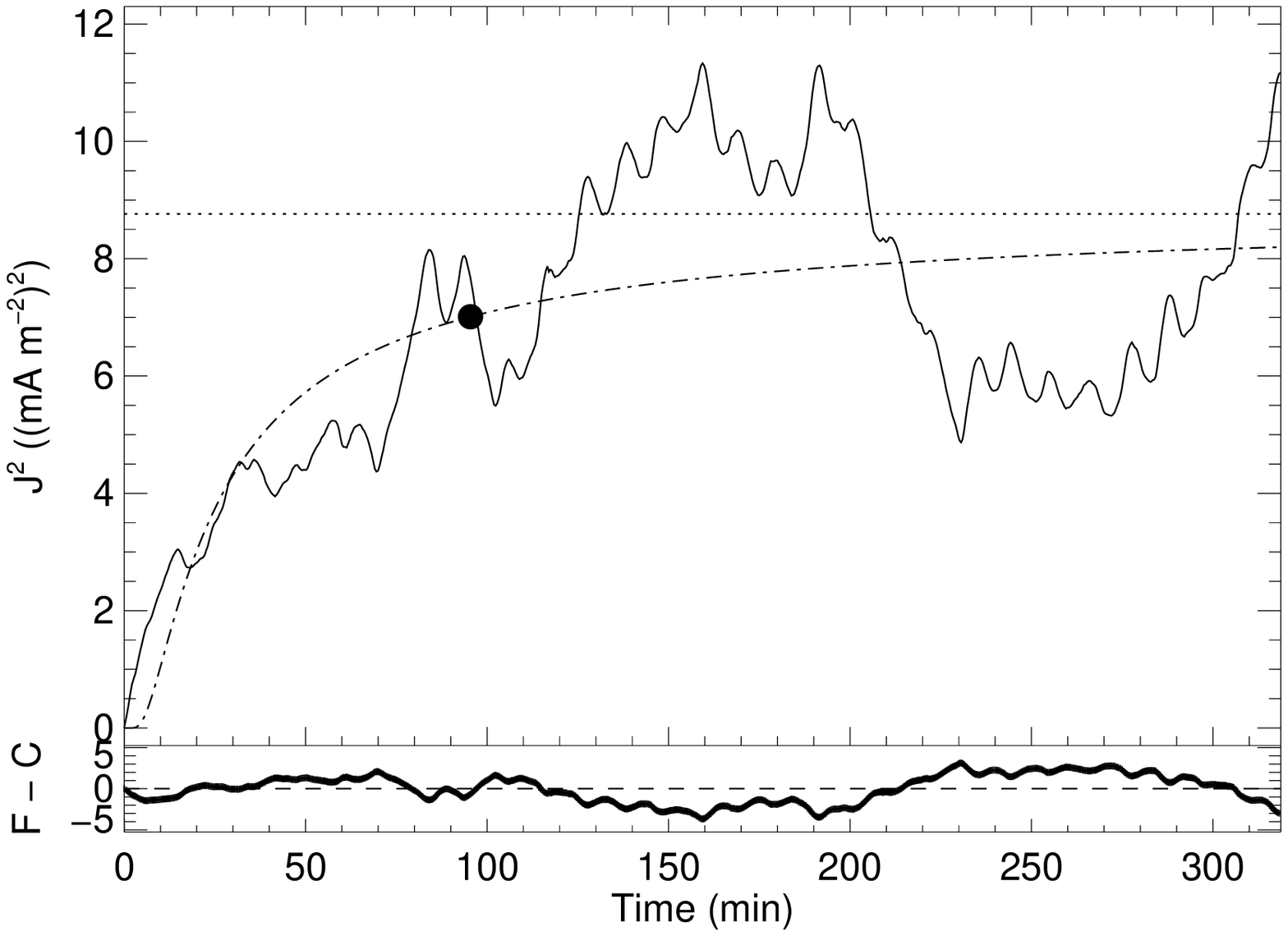}{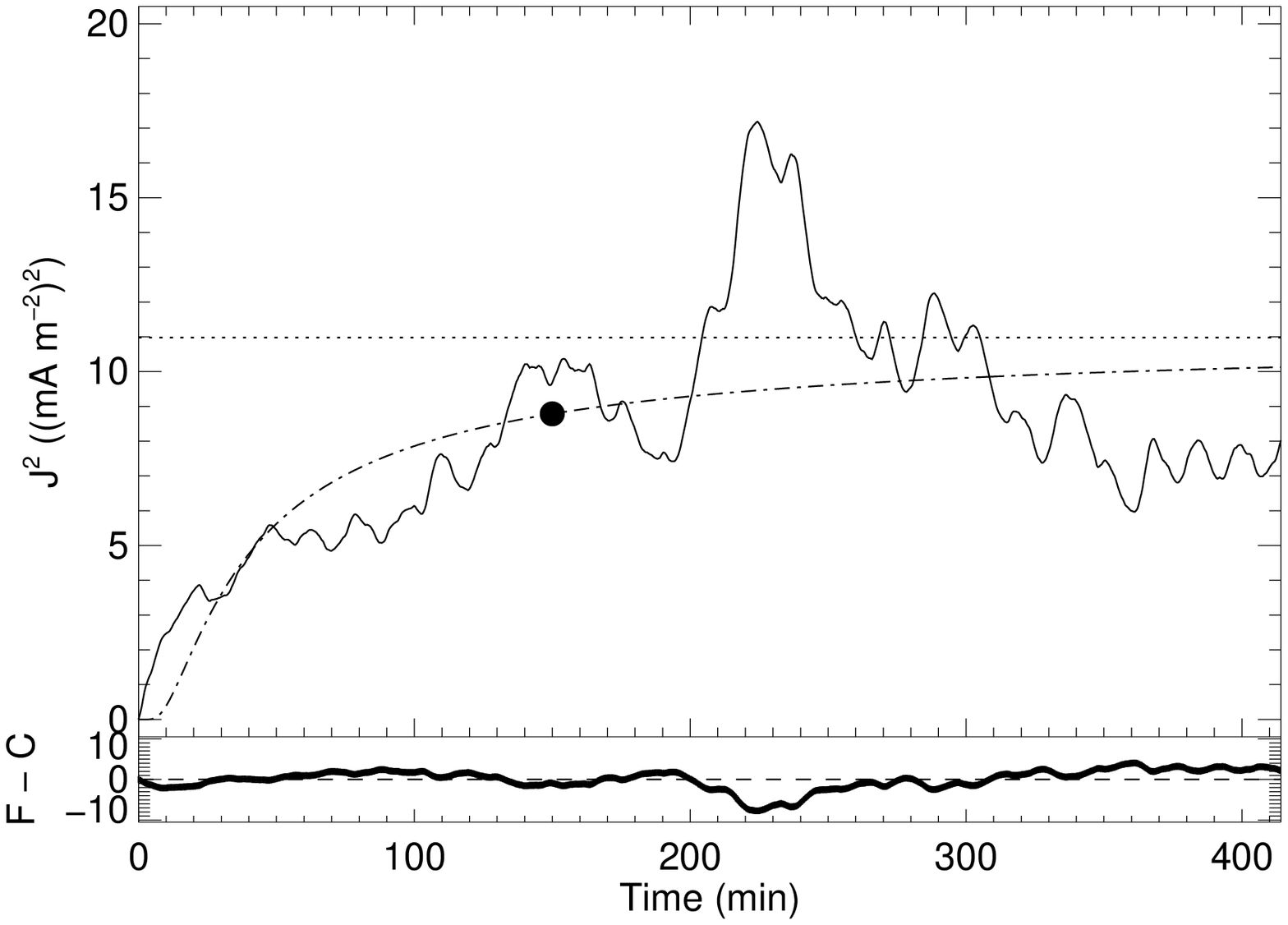}
\caption{Evolution of the average current density in the box as a function of time (solid line) and its curve fitting function (dash dotted line), in simulations IV (left) and V (right).\label{fig-45}}
\end{figure}
\clearpage

\begin{figure}
\epsscale{1.0}
\plottwo{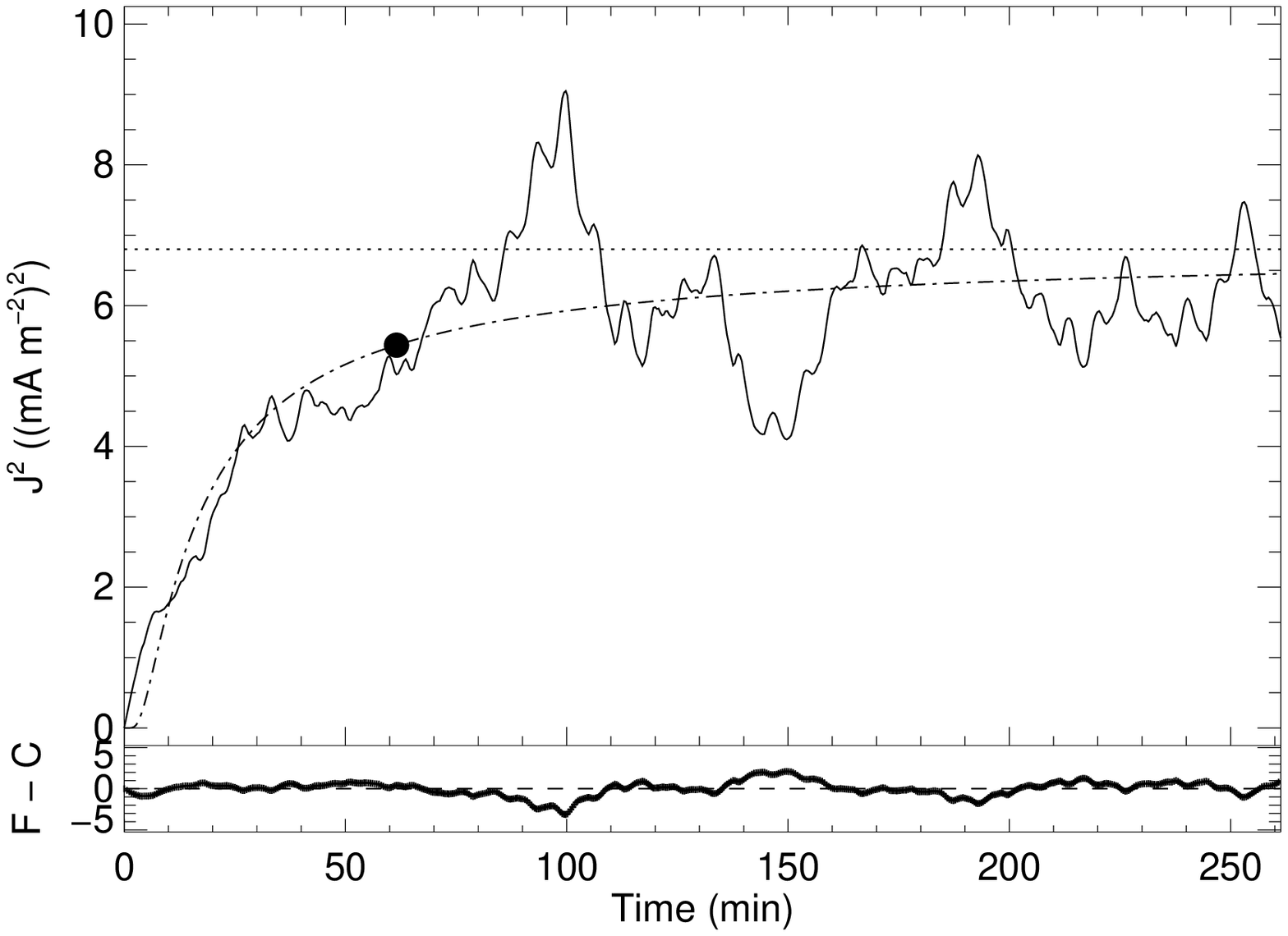}{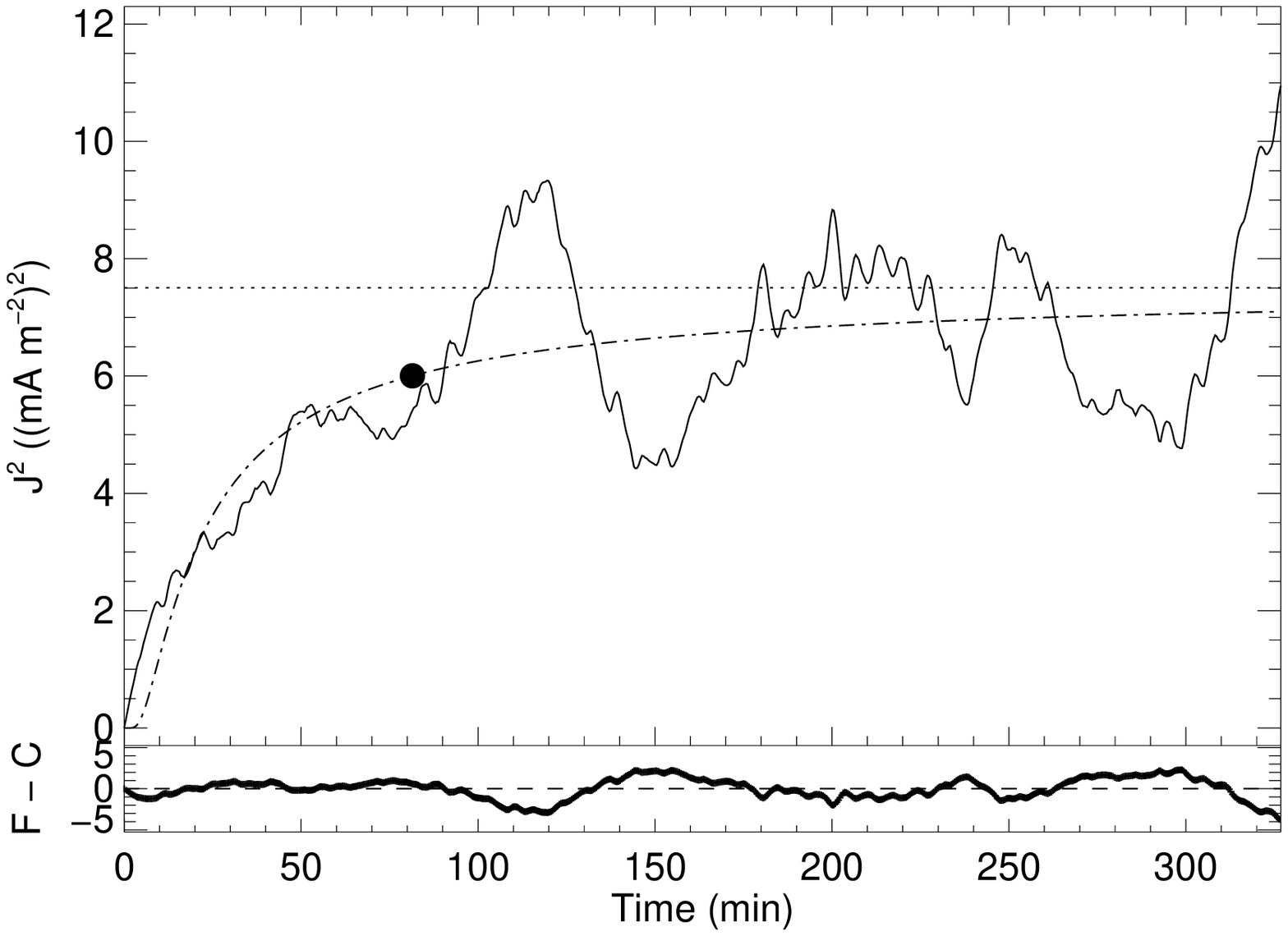}
\caption{Evolution of the average current density in the box as a function of time (solid line) and its curve fitting function (dash dotted line), in simulations VI (left) and VII (right).\label{fig-67}}
\end{figure}
\clearpage

\begin{figure}
\epsscale{1.0}
\plotone{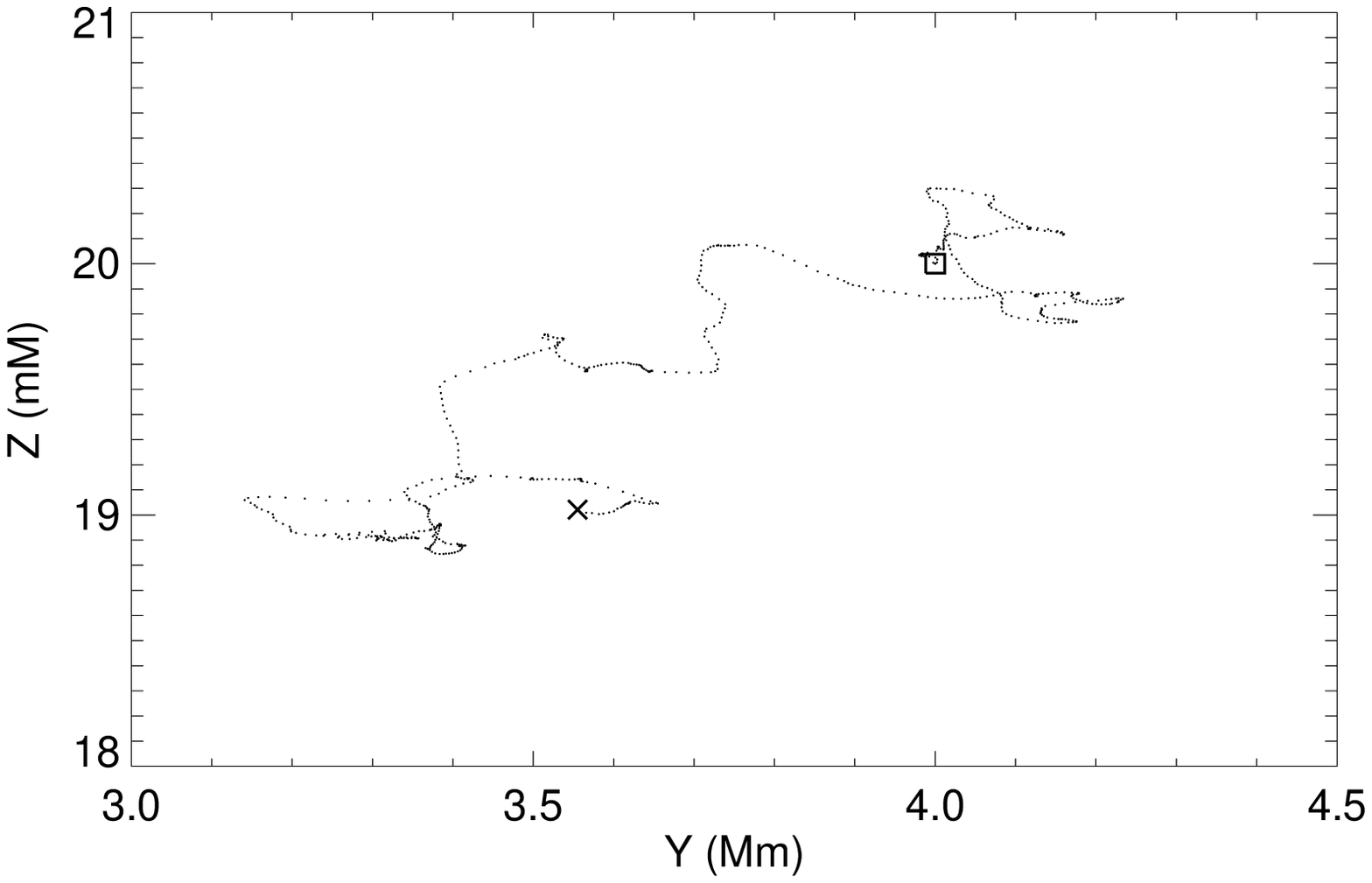}
\caption{Random motion of the footpoint of a magnetic field line, at the bottom of the box. The square symbol marks the initial position and the `X' shows the position of the footpoint at the end of the simulation I.\label{fig-8}}
\end{figure}
\clearpage

\begin{figure}
\epsscale{1.0}
\plotone{f9.ps}
\caption{Distance $R$ from the original location as a function of time for 1024 footpoints in simulation I. The dash dotted  line corresponds to the average distance and the solid line represents the fit assuming a fitting function $R=Ct^p$.\label{fig-9}}
\end{figure}
\clearpage

\begin{figure}
\epsscale{1.0}
\plotone{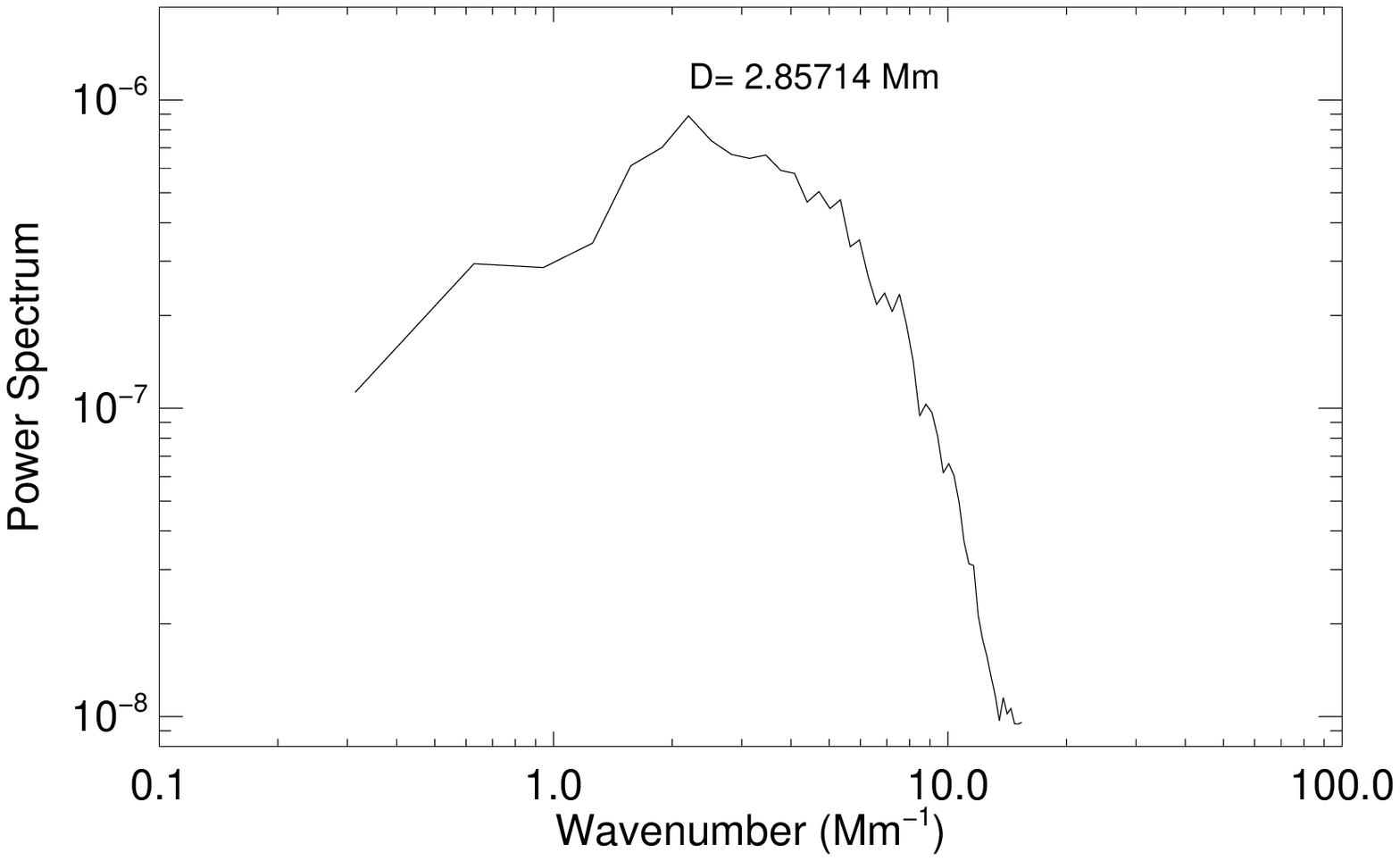}
\caption{Velocity power spectrum vs. wavenumber for a snapshot of simulation I. The figure shows the peak as a sign of the granulation pattern at the bottom of the box.\label{fig-10}}
\end{figure}
\clearpage

\begin{figure}
\epsscale{1.0}
\plotone{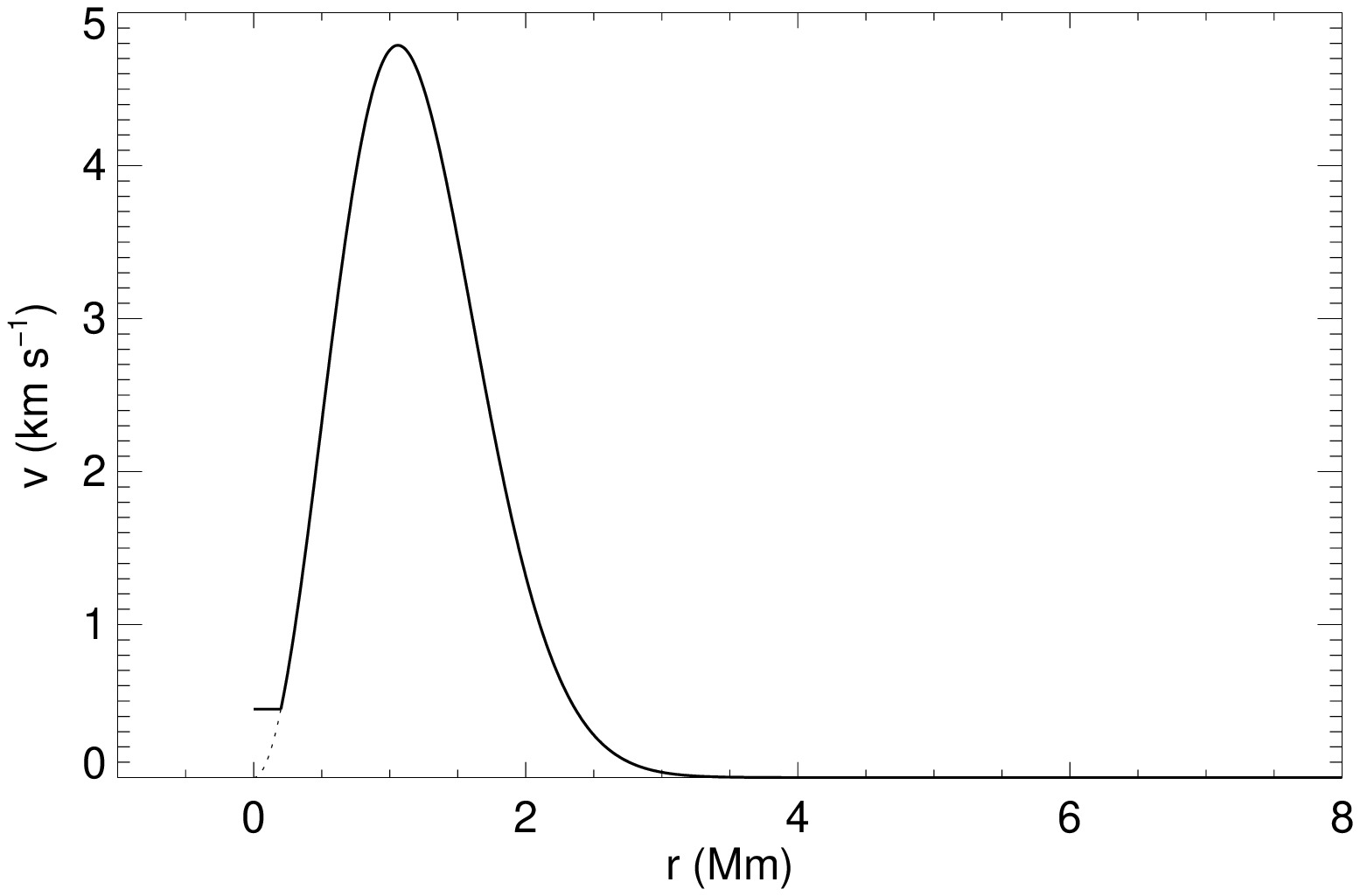}
\caption{Example of the velocity profile of a granule belonging to simulation I. The velocity can never be zero inside the code in order to keep the photospheric motions, so instead of the dashed portion of the plot a constant value of $v=0.2~\rm{km~s^{-1}}$ is assumed.\label{fig-11}}
\end{figure}
\clearpage

\begin{figure}
\epsscale{1.0}
\plotone{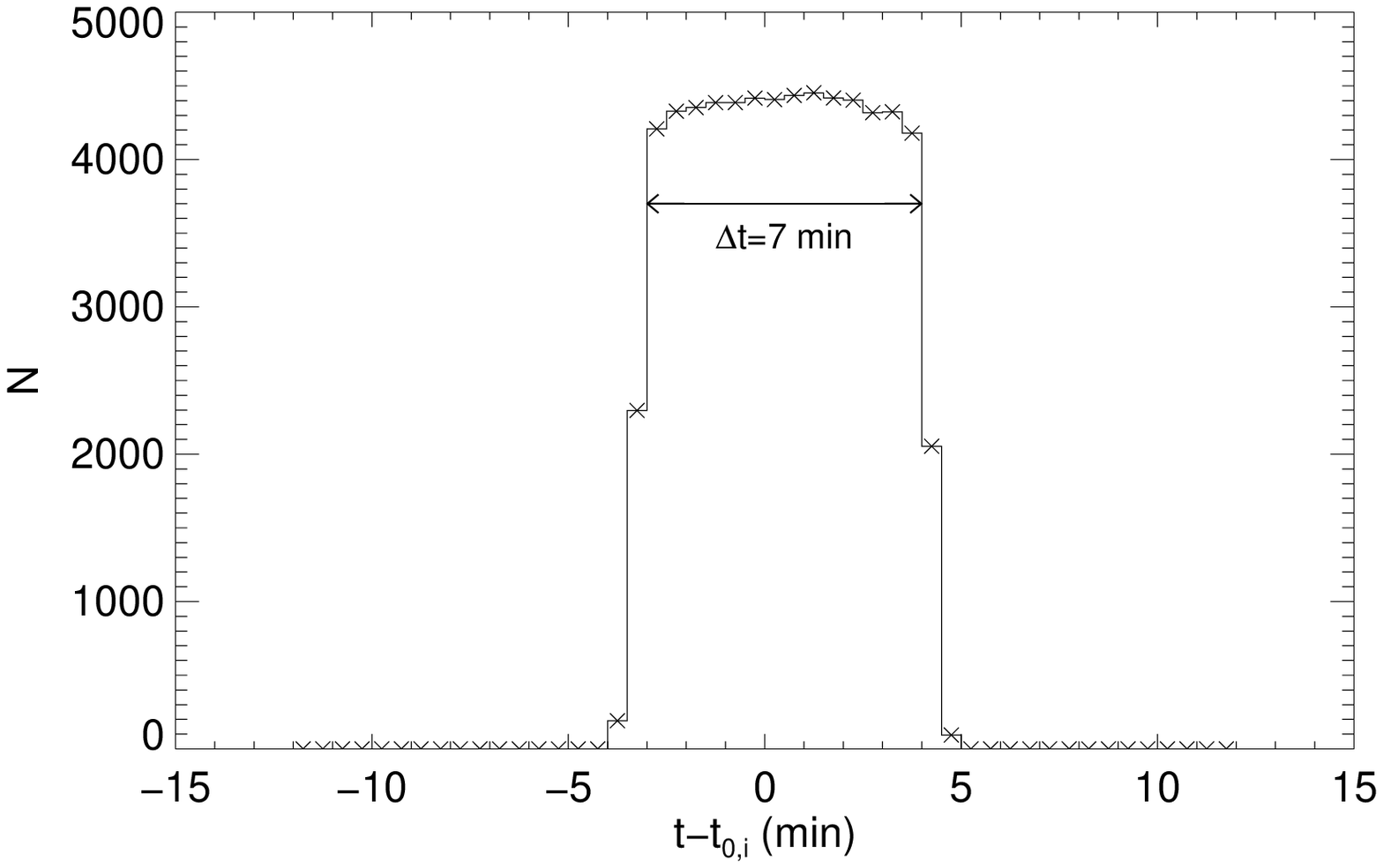}
\caption{Histogram showing the distribution of granules according to their development in simulation I. Negative values on the x-axis correspond to the period pre-full development while the positive values represent post-full development phase.\label{fig-12}}
\end{figure}
\clearpage

\begin{figure}
\epsscale{1.0}
\plottwo{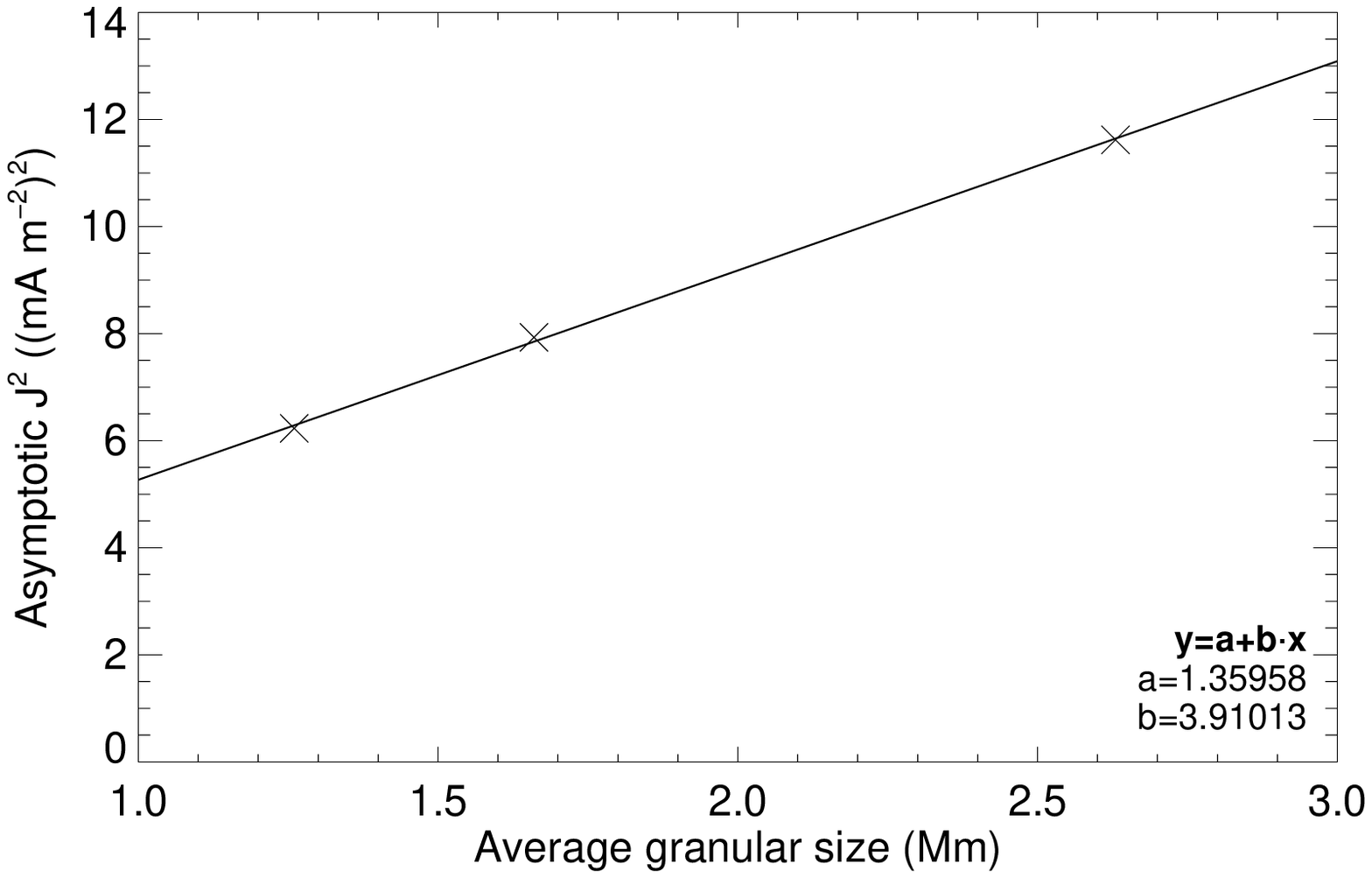}{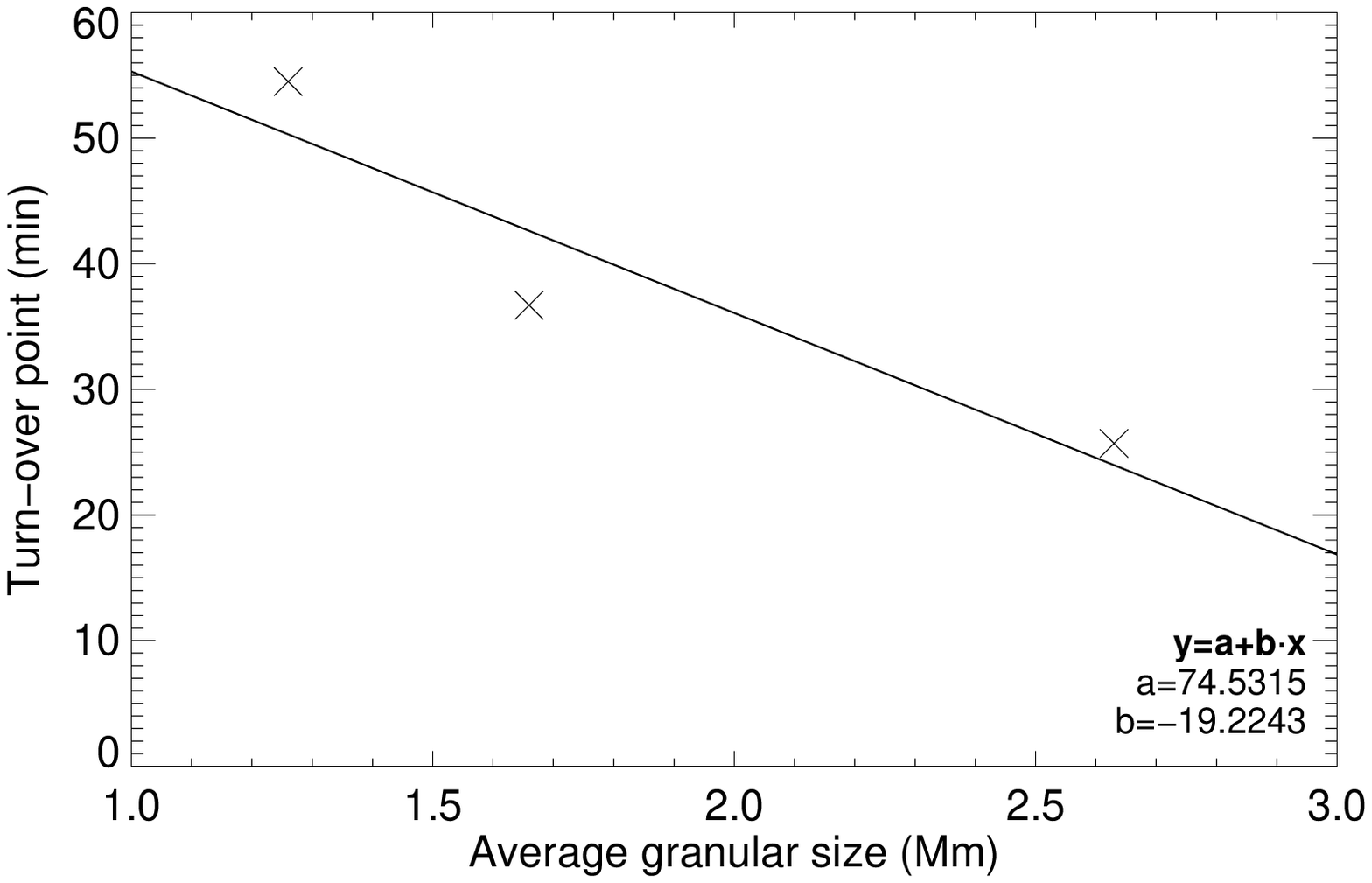}
\caption{$J_A^2$ (left) and turn-over points (right) as a function of the average granular size, from simulations of set 1.\label{fig-1314}}
\end{figure}
\clearpage

\begin{figure}
\epsscale{1.0}
\plottwo{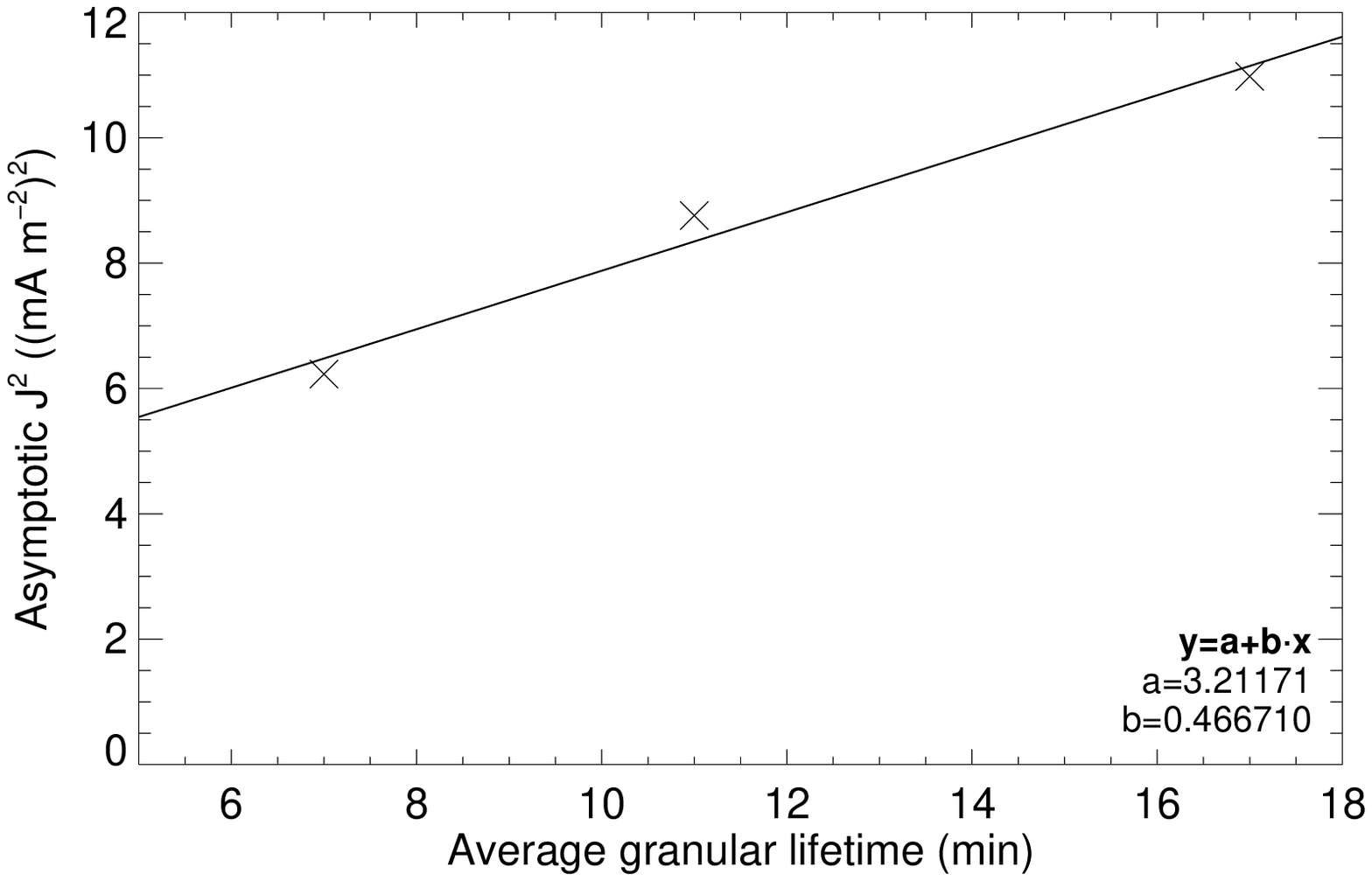}{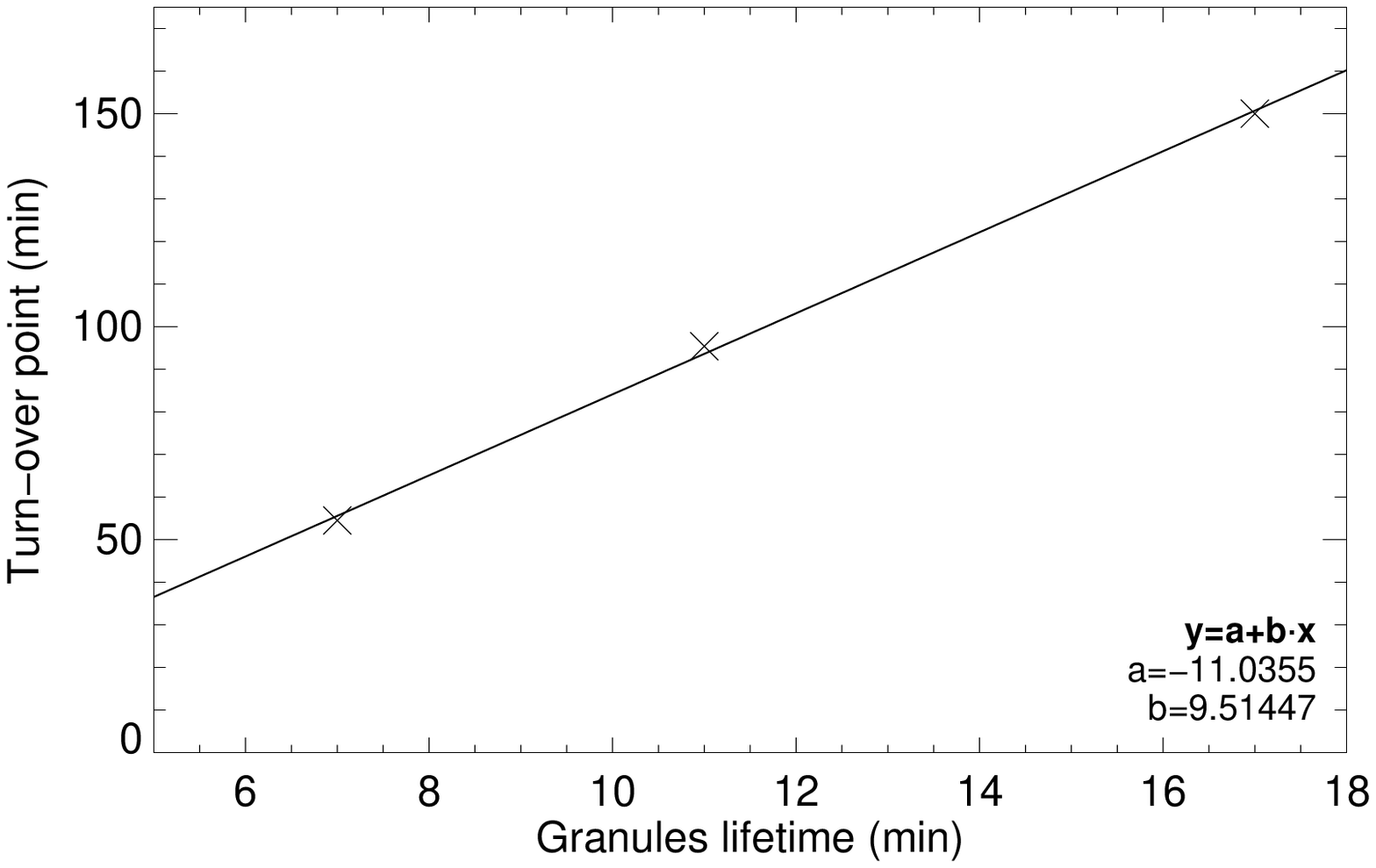}
\caption{$J_A^2$ (left) and turn-over points (right) as a function of the average granular lifetime, from simulations of set 2.\label{fig-1516}}
\end{figure}
\clearpage

\begin{figure}
\epsscale{1.0}
\plottwo{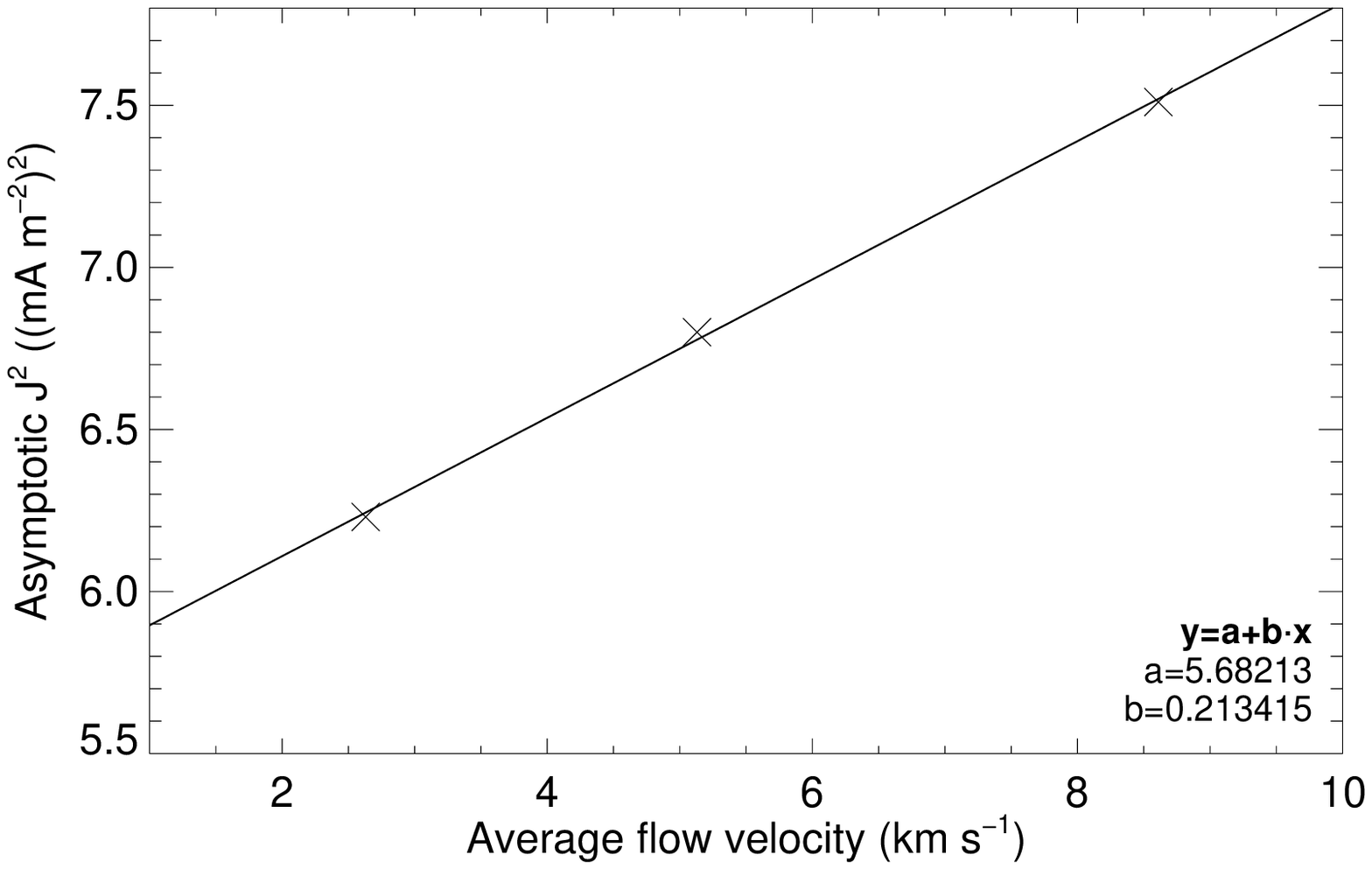}{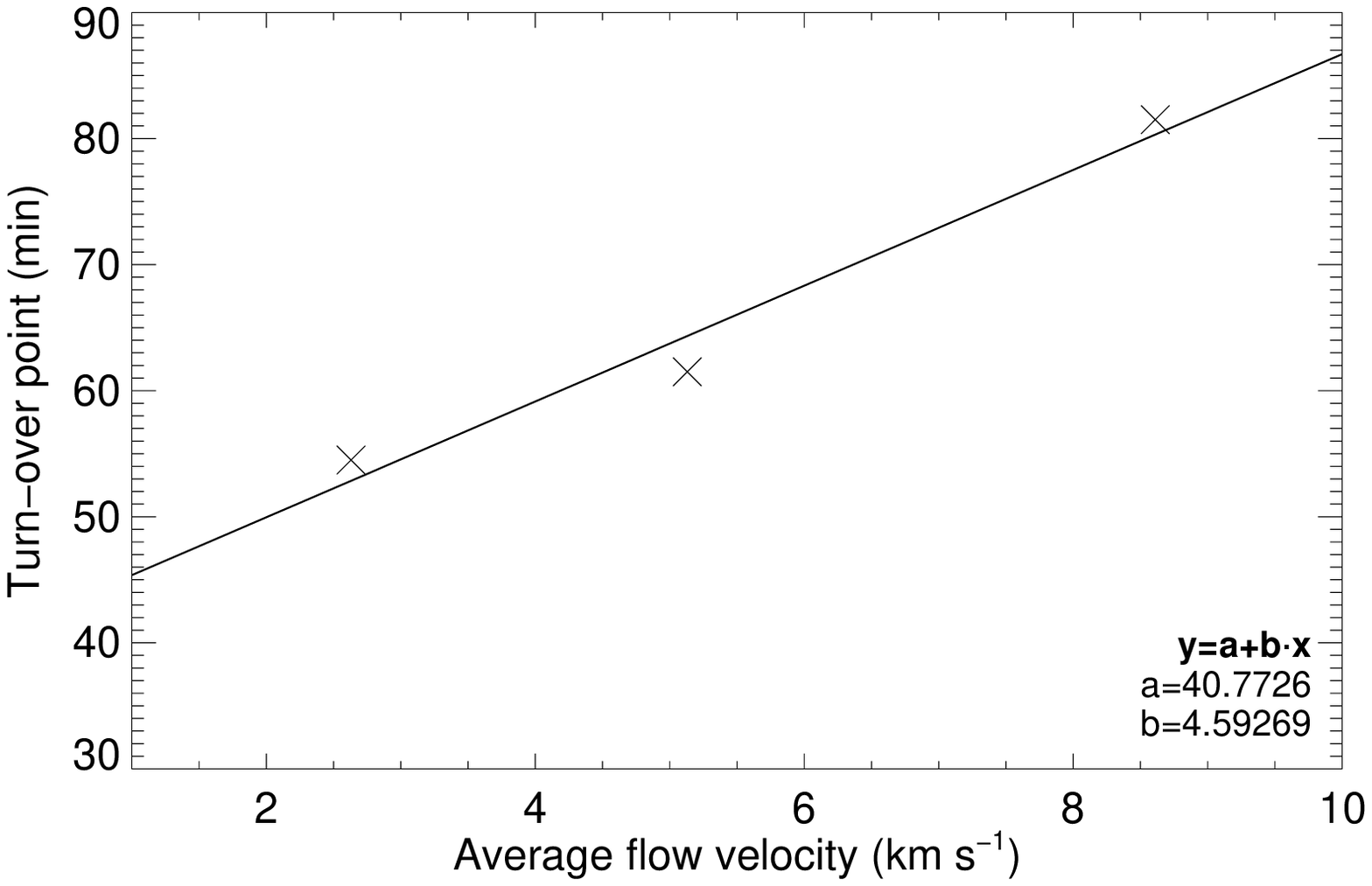}
\caption{$J_A^2$ (left) and turn-over points (right) as a function of the average flow velocity, from simulations of set 3.\label{fig-1718}}
\end{figure}
\clearpage

\begin{figure}
\epsscale{1.0}
\plottwo{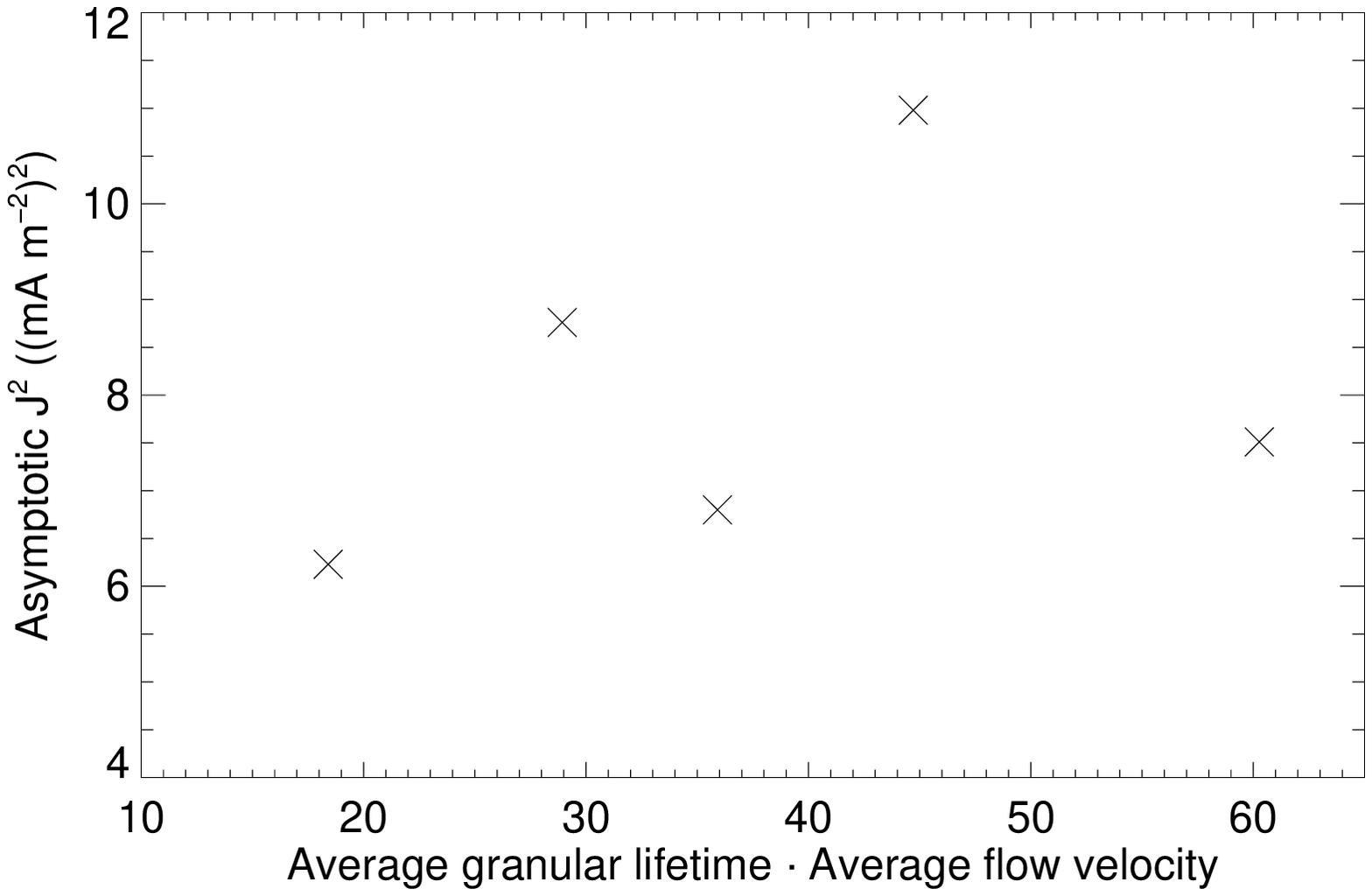}{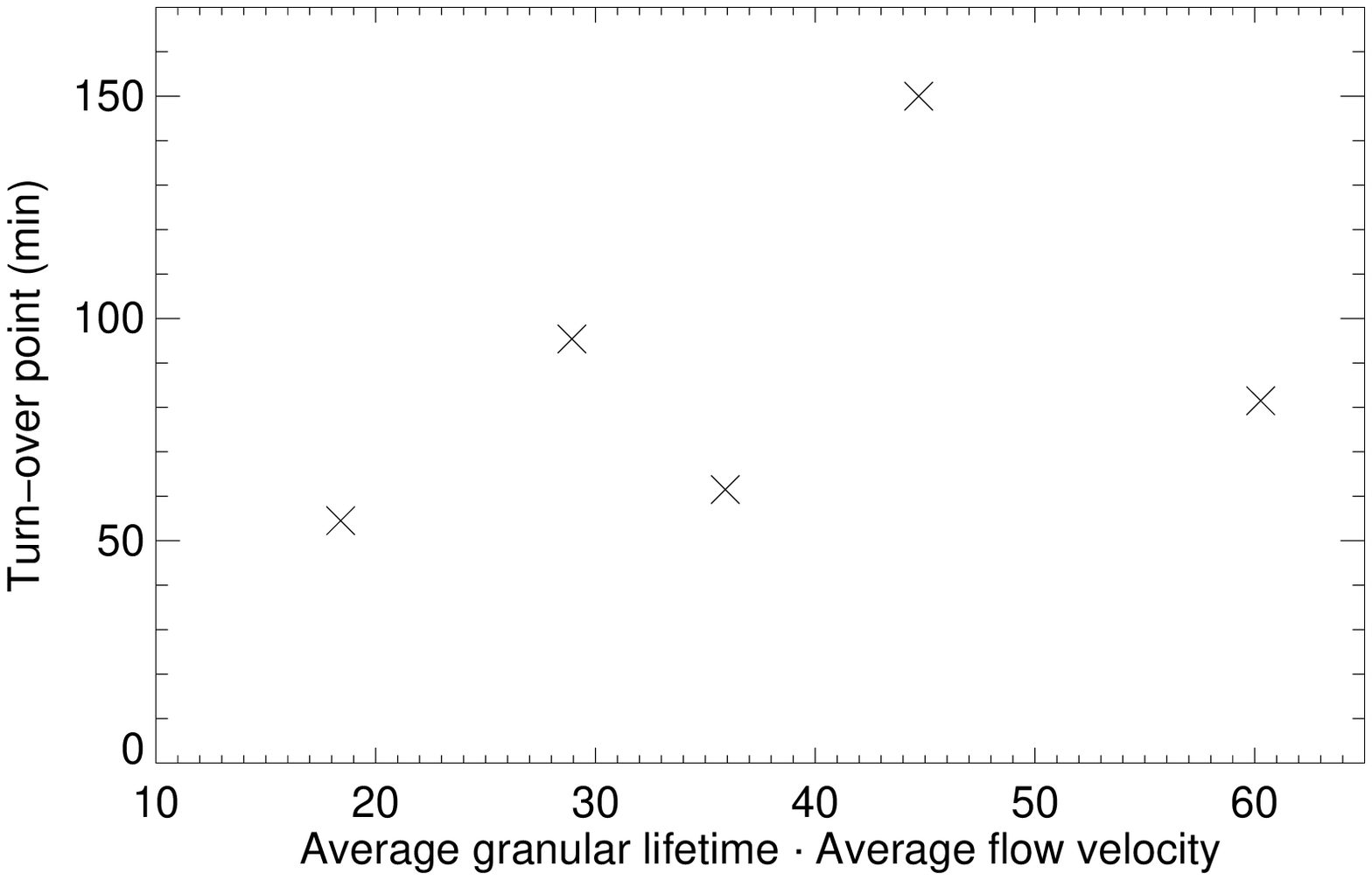}
\caption{$J_A^2$ (left) and turn-over points (right) as a function of the average granular lifetime $\times$ average flow velocity, from simulations I, IV, V, VI and VII.\label{fig-1920}}
\end{figure}
\clearpage

\fi

\ifnum\apjournal =1
\begin{deluxetable}{ccccccccc}
\tablecaption{Main parameters for the simulations \label{tab-1}}
\tablehead{
\colhead{} & \colhead{} & \multicolumn{3}{c}{Input parameters} & \colhead{} & \colhead{} & \colhead{}\\
\cline{3-5}
\colhead{Simulation} & \colhead{Set} & \colhead{$r$ (Mm)} & \colhead{$\tau$ (min)} & \colhead{$v~\rm{(km \cdot s^{-1}})$} & \colhead{\# Snapshots} & \colhead{$t$ (min)}
}
\startdata
I & 1, 2, 3 & 2 & 5 & 3 & 600 &  259.6\\
II & 1 & 3 & 5 & 3 & 600 & 251.3\\
III & 1 & 5 & 5 & 3 & 300 & 128.7\\
IV & 2 & 2 & 8 & 3 & 750 & 318.7\\
V & 2 & 2 & 12 & 3 & 1000 & 414.1\\
VI & 3 & 2 & 5 & 6 & 600 & 261.3\\
VII & 3 & 2 & 5 & 10 & 750 & 326.8\\
\enddata
\tablecomments{Variables on the table are $r$, the size of the granules; $\tau$, the lifetime of the granules; $v_{0}$, the plasma flow velocity, $t$, the time-span of the simualtion.}
\end{deluxetable}
\clearpage

\begin{deluxetable}{cccccc}
\tablecaption{Fitting parameters ($a_{0}$ and $a_{1}$) in Eq. \ref{eq:fit}, their uncertainties ($\epsilon_{a_0}$ and $\epsilon_{a_1}$) and turn-over points. \label{tab-2}}
\tablehead{
\colhead{} & \colhead{} & \multicolumn{4}{c}{Fitting-function}\\
\cline{3-6}
\colhead{Simulation} & \colhead{Turn-over point (min)} & \colhead{$a_{0}$} & \colhead{$\epsilon_{a_0}$} & \colhead{$a_{1}$} & \colhead{$\epsilon_{a_1}$}
}
\startdata
I & 54.5 & 6.23 & 0.06 & 12.2 & 0.7\\
II & 36.7 & 7.93 & 0.08 & 8.15 & 0.61\\
III & 25.7 & 11.6 & 0.2 & 5.78 & 0.49\\
IV & 95.4 & 8.76 & 0.11 & 21.2 & 1.3\\
V & 150.0 & 10.97 & 0.15 & 33.2 & 2.0\\
VI & 61.5 & 6.80 & 0.06 & 13.8 & 0.7\\
VII & 81.5 & 7.50 & 0.08 & 18.2 & 1.0\\
\enddata
\end{deluxetable}
\clearpage

\begin{deluxetable}{cccccc}
\tablecaption{Values of $p$ and uncertainty $\epsilon_{p}$ in Eq. \ref{eq:randomwalk} for each simulation \label{tab-3}}
\tablehead{
\colhead{Simulation} & \colhead{p} & \colhead{$\epsilon_{p}$}
}
\startdata
I & 0.630 & 0.007\\
II & 0.311 & 0.005\\
III & 0.610 & 0.013\\
IV & 0.547 & 0.006\\
V & 0.530 & 0.005\\
VI & 0.585 & 0.009\\
VII & 0.623 & 0.005\\
\enddata
\end{deluxetable}
\clearpage

\begin{deluxetable}{cccccc}
\tablecaption{Average granular size ($r_{av}$), step length ($\ell$), average velocity ($\bar v$), average lifetime ($\bar \tau$) and average number of random steps ($N$) for each experiment \label{tab-4}}
\tablehead{
\colhead{Simulation} & \colhead{$r_{av}$ (Mm)} & \colhead{$\ell$ (Mm)} & \colhead{$\bar v~\rm{(km \cdot s^{-1})}$} & \colhead{$\bar \tau~\rm{(min)}$} & \colhead{N}
}
\startdata
I & 1.26 & 1.10 & 2.63 & 7 & 37.1\\
II & 1.66 & 1.10 & 2.29 & 8 & 31.4\\
III & 2.63 & 1.11 & 2.17 & 8.5 & 15.1\\
IV & 1.32 & 1.32 & 2.58 & 11 & 29.0\\
V & 1.29 & 1.29 & 2.63 & 17 & 24.4\\
VI & 1.25 & 1.25 & 5.13 & 7 & 37.3\\
VII & 1.30 & 1.30 & 8.61 & 7 & 46.7\\
\enddata
\end{deluxetable}
\clearpage

\begin{deluxetable}{cccccc}
\tablecaption{Values of the average displacement at the turn-over point ($R_{turn}$) and estimates for the critical angle($\theta_{est}$) \label{tab-5}}
\tablehead{
\colhead{Simulation} & \colhead{$R_{turn}$ (Mm)} & \colhead{$\theta_{est}$}
}
\startdata
I & 3.08 & 8.76$^\circ$\\
II & 2.35 & 6.71$^\circ$\\
III & 1.92 & 5.50$^\circ$\\
IV & 3.89 & 11.0$^\circ$\\
V & 3.83 & 10.9$^\circ$\\
VI & 3.71 & 10.5$^\circ$\\
VII & 4.44 & 12.5$^\circ$\\
\enddata
\end{deluxetable}
\fi

\end{document}